\documentclass[%
 reprint,
superscriptaddress,
 amsmath,amssymb,
prb,
]{revtex4-2}

\usepackage{graphicx}% Include figure files
\usepackage{dcolumn}% Align table columns on decimal point
\usepackage{bm}% bold math
\usepackage[bookmarks=true,colorlinks=true,urlcolor=blue,linkcolor=blue,citecolor=blue,breaklinks=true]{hyperref}
\PassOptionsToPackage{hyphens}{url}
\usepackage{graphicx}
\usepackage{amsmath, amssymb}
\usepackage[table]{xcolor}       %  provides easy access to several kinds of color tints, shades, tones, and mixes of arbitrary colors

\usepackage{lipsum}

\everymath{\medmuskip=1.5mu minus 1.5mu\thickmuskip=2mu minus 2mu}

\newcommand{\DIPC}[0]{
Donostia International Physics Center (DIPC),
Paseo Manuel de Lardizabal 4, 20018 Donostia-San Sebasti\'an, Spain}
\newcommand{\CFM}[0]{
Centro de F\'{\i}sica de Materiales CFM/MPC (CSIC-UPV/EHU), Paseo Manuel de Lardizabal 5, 20018 Donostia-San Sebasti\'an, Spain}
\newcommand{\PolymerEHU}[0]{Departamento de Pol\'{i}meros y Materiales Avanzados: F\'{i}sica, Qu\'{i}mica y Tecnolog\'{i}a, Facultad de Qu\'{i}micas (UPV/EHU), Apartado 1072, 20080 Donostia-San Sebasti\'{a}n, Spain}

\newcommand{\UPC}[0]{Departament de F\'{i}sica, Universitat Polit\`{e}cnica de Catalunya, Campus Nord B4-B5, E-08034, Barcelona, Spain}

\newcommand{\IFS}[0]{Centre for Advanced Laser Techniques, Institute of Physics, Bijeni\v{c}ka 46, 10000 Zagreb, Croatia}

\begin{document}

%\preprint{APS/123-QED}

%\title{\textcolor{blue}{Vibrational dynamics of CO on Pd(111) under thermal and non-equilibrium conditions}}
\title{Vibrational dynamics of CO on Pd(111) in and out of thermal equilibrium}
%\title{Thermal and non-equilibrium vibrational relaxation of CO on Pd(111)}% Force line breaks with \\
\author{Raúl Bombín}
\email{raul.bombin@ehu.eus}
\affiliation{\PolymerEHU}
\affiliation{\CFM}
\affiliation{\UPC}

\author{A. S. Muzas}
\affiliation{\CFM}
\author{Dino Novko}
\affiliation{\IFS}
%\affiliation{\DIPC}
%
\author{J. I\~{n}aki Juaristi}
\email{josebainaki.juaristi@ehu.eus}
\affiliation{\PolymerEHU}
\affiliation{\CFM}
\affiliation{\DIPC}
\author{Maite Alducin}
\email{maite.alducin@ehu.eus}
\affiliation{\CFM}
\affiliation{\DIPC}

\date{\today}

\begin{abstract}
Using many-body perturbation theory and density functional perturbation theory, we study the vibrational spectra of the internal stretch (IS) mode of CO on Pd(111) for the bridge and hollow adsorption structures that are experimentally identified at 0.5~ML coverage. Our theoretical treatment allows us to determine the temperature dependence of the IS vibrational spectra under thermal conditions as well as the time evolution of the non-equilibrium transient spectra induced by femtosecond laser pulses. Under thermal conditions (i.e., for equal electronic $T_e$ and phononic $T_l$ temperatures),
the calculated lifetimes at 10-150~K 
are mostly due to nonadiabatic couplings (NC), i.e., first-order electronic excitations. As temperature increases, also the contribution of the second-order electron mediated phonon-phonon couplings (EMPPC) progressively increases from 25\% at low temperatures to 50\% at 300~K. Our calculations for the laser-induced non-equilibrium conditions comprise experimental absorbed fluences of 6-130~J/m$^2$. For fluences for which $T_e>$2000~K,  
the transient vibrational spectra are characterized by two different regimes that follow the distinct time-evolution of $T_e$ and $T_l$ and are respectively dominated by NC and EMPPC processes. At lower fluences, the initial fast regime becomes progressively negligible as $T_e$ decreases and only the steady second regime remains visible. Qualitatively, all these spectral properties are common to the both adsorption structures studied here.
\end{abstract}

\maketitle

%\tableofcontents
\section{Introduction}

The internal stretch (IS) mode  of polar molecules adsorbed on metal surfaces, being decoupled in energy from the rest of vibrational modes of the system, can be monitored during a reaction in time-resolved vibrational
spectroscopy experiments~\cite{arnolds2010}. In these kind of setups, the adsorbate dynamics is initiated by a pump laser pulse and the frequency shift and the linewidth of the IS mode are tracked with femtosecond time resolution. It has been suggested that these changes can provide insight into the initial steps of the induced reaction~\cite{Germer93,Bonn2000,backus05, Inoue2016}, including the transient charge flow that the pump laser may induced in organic adsorbates~\cite{GarciaRey17,GarciaRey19}.

Commonly, after the pump pulse heats the surface, the IS mode exhibits a rapid initial redshift followed by a slower (but also fast) recovery (e.g., CO/Ru(0001)~\cite{Bonn2000}, NO/Ir(111)\cite{Lane2006}, CO/Ir(111)~\cite{Lane2007}, CO/Pt(111)~\cite{Beckerle91,Fournier2004,Watanabe2010,Inoue2012}, CO/Cu(100)~\cite{Germer94,Inoue2016}, CO/Cu(110)~\cite{Omiya2014}). However, for the case of CO molecules coordinated to ruthenium tetraphenylporphyrin on a Cu(110) surface a blueshift has been reported~\cite{Omiya2019}.
Closely related, thermal heating experiments represent another example in which an increase of the linewidth and a frequency redshift in the IS mode is expected as temperature rises. The latter has been observed for CO/Cu(100)~\cite{Germer94,Cook1997,cook97,Novko2018}, CO/Pt(111)\,\cite{Schweizer89,Beckerle91,Carrasco10}, CO/Ru(0001)\,\cite{Hoffmann86}, CO/Ru(0101)\,\cite{Symonds04}, as well as for CO/Ir(111) and NO/Ir(111)~\cite{Lane2007}. 
However, Persson \textit{et al.} reported a thermal blueshift for CO/Ni(111)~\cite{Person1985} when varying the temperature from 25 to 300~K. Under similar temperature changes, a blueshift has been also observed in a high coverage phase, with a rather more complex structure, of CO/Pd(100)~\cite{Cook1997}. All these findings suggest that an accurate microscopic description of the different mechanisms that participate in the vibrational dynamics of polar molecules on the different metallic surfaces is needed in order to give a correct interpretation to experimental measurements. 

In our previous work we studied the frequency shift that is induced on the IS mode of the 0.5~ML of CO on a Pd(111) surface under pump-probe conditions~\cite{Bombin2023}. There, we showed that due to the sharp shape of the electron density of states (DOS) around the Fermi level, the electron-phonon (\textit{e}-ph) coupling is effectively screened~\cite{Bombin2023}, giving place to an anomalous transient blueshift in the initial few hundreds of femtoseconds upon arrival of the pump laser pulse. The time evolution of the observed blueshift resembles that of Ref.~\cite{Omiya2019}, what may hint at a similar mechanism behind their origin. Here we extend that study to the whole range of thermal and pump-probe non-equilibrium experimental conditions, characterizing not only the frequency-shift but also the changes that are induced in the linewidth. As there has been a long debate about the adsorption structure of the 0.5~ML of CO on Pd(111), we study both the fcc-hcp and bridge structures that have been experimentally observed~\cite{Rose2002,Surnev2000}. All the calculations presented here are performed with the methodology of Refs.~\cite{Novko2018,Novko2019}, which is based on many-body perturbation theory (MBPT) and density functional perturbation theory (DFPT). This robust theoretical framework has been successfully applied to explain the transient subpicosecond vibrational spectral changes measured for  CO/Cu(100)~\cite{Novko2018,Novko2019}.

The paper is organized as follows. In Sec.~\ref{sec:methods} we describe the method that we use to evaluate nonadiabatic effects in the IS vibrational mode of the CO adlayer. In the MBPT framework this is done in terms of the phonon self-energy. In Sec.~\ref{sec:system} we characterize the electronic and phononic properties of the system under study. Further details for the Pd(111) surface and CO/Pd(111) system that are not essential for the discussion presented in the main text are provided in Appendices~\ref{app:Pdslab} and \ref{app:COonPd}. Our results for the frequency shift and linewidth of the CO IS mode under thermal and non-equilibrium conditions are shown in  Secs.~\ref{sec:thermal} and~\ref{sec:nothermal}, respectively. A comparison to other theoretical and experimental studies on the dynamics of the IS mode of polar molecules in other transition metal surfaces is presented in Sec.\ref{sec:comparison}. Finally, the main conclusions are summarized in Sec.~\ref{sec:conclusions}.

\section{Methods}
\label{sec:methods}
Electronic structures, dynamical matrices, and phonon perturbation potentials are calculated with density functional theory (DFT) and DFPT using the \textsc{quantum espresso} first principles package \cite{Giannozzi2009,Giannozzi2017}. The exchange-correlation interaction is described with the Bayesian error estimation functional with van
der Waals correlation (BEEF-vdW)~\cite{Wellendorff2012}, which is specifically designed for treating surface science problems. We use the ultrasoft pseudopotentials by A. dal Corso~\cite{dalcorso2014} with a kinetic energy cutoff of 1360 eV and a Gaussian smearing of 0.27~eV for the electronic state occupancies. The electronic states and charge densities are evaluated by sampling the Brillouin zone with a $\Gamma$-centered 12$\times$12$\times$1 Monkhorst-Pack (MP) mesh~\cite{Monkhorst1976}. The CO/Pd(111) system is modeled with a $c(2\times \sqrt{3})$rect surface unit cell that includes four Pd-layers with 12.9~{\AA} of vacuum. The desired 0.5~ML of CO coverage is achieved by including two CO molecules in the supercell. All the atoms in the supercell are relaxed until the Hellmann-Feynman forces on each atom are below 2$\times$10$^{-5}~$~Ry/$a_\mathrm{B}$ ($\sim$2$\times$10$^{-4}$~eV/\AA). See Appendices~\ref{app:Pdslab} and \ref{app:COonPd} for more details about these structures. The dynamical matrices and perturbation potentials are calculated on a $\Gamma$-centered 6$\times$6$\times$1 $\textbf q$-grid. 

We use the Electron-phonon Wannier (\textsc{epw}) code to evaluate the \textit{e}-ph matrix elements~\cite{NOFFSINGER20102140,PONCE2016116}. From the electronic states and charge density computed in the $\Gamma$-centered $\textbf k$-grid we construct a set of 108 maximally localized Wannier functions (MLWFs), using the selected columns of density matrix procedure~\cite{Vitale2020}. The \textit{e}-ph matrix elements are calculated in the $\textbf k$ and $\textbf q$ grids that have been used  to sample the electronic and phononic Brillouin zones. Finally, taking advantage of the  properties of the MLWFs~\cite{Mostofi2008,Marzari2012,Mostofi2014,Pizzi2020}, the \textit{e}-ph matrix elements are interpolated on finer $\textbf k$- and $\textbf q$-grids to obtain the desired accuracy in the properties that we study.

The vibrational spectra of CO on Pd(111) are studied within MBPT by calculating the phonon self-energy due to \textit{e}-ph coupling $\pi_\lambda(\textbf{q},\omega_{\textbf{q},\lambda})$ ($\lambda$, $\textbf{q}$, and $\omega_{\textbf{q},\lambda}$ are the index, momentum, and energy of the phonon mode, respectively). The corresponding phonon linewidth is determined by taking the imaginary part of this quantity as, $\gamma_{\textbf{q},\lambda} = -2\mathrm{Im}\pi_\lambda(\textbf{q},\omega_{\textbf{q},\lambda})$, while the renormalization of the phonon frequency  
is obtained from the real part as,  $\omega_{\textbf{q},\lambda}^2 = \omega^2_\mathrm{A} + 2\omega_\mathrm{A}\mathrm{Re}\left[\pi_\lambda(\textbf{q},\omega_{\textbf{q},\lambda})-\pi_\lambda(\textbf{q},0)\right]$, with $\omega_\mathrm{A}$ being the adiabatic phonon frequency obtained in the DFPT calculation~\cite{Novko2016,Giustino2017}. As nonadiabatic corrections to the phonon energy are usually small compared to the adiabatic energy (i.e., $\omega_{\textbf{q},\lambda}-\omega_\mathrm{A} \ll\omega_\mathrm{A}$), the renormalized phonon frequency can be approximated by $\omega_{\textbf{q},\lambda} \approx \omega_\mathrm{A} + \mathrm{Re}\left[\pi_\lambda(\textbf{q},\omega_{\textbf{q},\lambda})-\pi_\lambda(\textbf{q},0)\right]$. 
Here we are interested in the vibrational spectra probed by infrared light. Thus, only the long wavelength part of the phonon self-energy, denoted hereafter as $\pi_{\lambda}(\omega_{0,\lambda})\equiv\pi_{\lambda}(\mathbf{q}\approx0,\omega)$, will be taken into account.

Following previous works~\cite{Novko2018,Novko2019}, we consider that the phonon self-energy consists of first- and second-order terms in the \textit{e}-ph coupling, $\pi_\lambda(\omega_{0,\lambda}) = \pi_\lambda^{[1]}(\omega_{0,\lambda}) + \pi_\lambda^{[2]}(\omega_{0,\lambda})$. Later it will be clear that $\pi_\lambda^{[1]}$ and $\pi_\lambda^{[2]}$ correspond to dominant interband and intraband contributions, respectively. The first order accounts for nonadiabatic coupling (NC) to electron-hole pairs, while the second order corresponds to the so-called electron mediated phonon-phonon coupling (EMPPC). This theory has been able to reproduce and explain the thermal~\cite{Novko2018} and ultrafast~\cite{Novko2019} vibrational spectra of CO/Cu(100) measured in Refs.~[\onlinecite{Germer94}] and [\onlinecite{Inoue2016}], respectively.

\textit{Nonadiabatic coupling}.---  
The expression for the first-order NC term that exclusively accounts for the electron-hole pair (de)excitations reads~\cite{Novko2016, Giustino2017, Novko2018}
\begin{equation}
\pi_\lambda^{[1]}(\omega_{0,\lambda}; T_e)=\sum_{\mu \mu^\prime\textbf{k}\sigma}\left|g^{\mu\mu^\prime}_\lambda(\textbf{k},0)\right|^2\frac{f(\epsilon_{\mu\textbf{k}})-f(\epsilon_{\mu^\prime\textbf{k}})}{\omega_{0,\lambda} + \epsilon_{\mu\textbf{k}}-\epsilon_{\mu^\prime\textbf{k}}+ i\eta},
\label{eq:first_order}
\end{equation}
where $\mu$, $\textbf{k}$, and $\epsilon_{\mu\textbf{k}}$ are the electron band index, momentum, and energy, respectively; $g^{\mu\mu^\prime}_\lambda(\textbf{k},\textbf{q})$ are the \textit{e}-ph matrix elements; and the summation over $\sigma$ accounts for the spin degree of freedom. We have introduced the Fermi-Dirac distribution function $f(\epsilon_{\mu,\textbf{k}})=1/(e^{\beta[\epsilon_{\mu\textbf{k}}-\mu(T_e)]}+1)$, where $\beta=1/(k_BT_e)$, $k_B$ is the Boltzmann constant, $T_e$ is the electronic temperature, and $\mu(T_e)$ is the chemical potential that equals the Fermi level at $T_e = 0$, i.e.,  $\varepsilon_F=\mu(0)$. The chemical potential, is calculated self-consistently by solving numerically the equality,  $N_e = \int \mathrm{DOS}(\epsilon)f(\epsilon,T_e,\mu(T_e)) d\epsilon$, that assures conservation of the number of electrons $N_e$. Strictly speaking $\eta$ is an infinitesimal parameter and one obtains the exact first-order phonon linewidth in the limit $\eta\to 0$. However, $\eta$ is usually fixed to a small value representing the broadening of the electronic states due to scattering processes~\cite{Novko2016,Rittmeyer17}. In our case we fix it to a small, physically motivated value of 30~meV~\cite{Hayashi2013,Schendel2017}. Note that the terms $\mu = \mu^\prime$ vanish so that $\pi_\lambda^{[1]}$ only contains interband contributions. As a final remark, note that the expression for the linewidth derived from Eq.~\eqref{eq:first_order} is equivalent to that obtained from the Fermi golden rule that has been used by other authors~\cite{hellsing84,Sorbello85, Persson1984, Krishna2006, forsblom07, Askerka2016, maurer16, Diesen21, Schreck22}.

\textit{Electron mediated phonon-phonon coupling}.--- Early studies on the vibrational relaxation of polar molecules on metal surfaces already accounted for two-phonon processes that involved anharmonic coupling between the adsorbate vibrational modes~\cite{Persson1984}. However, the fact that the IS mode is decoupled in energy from the other vibrational modes makes this mechanism quite inefficient. Nonetheless, as shown in Ref.~\cite{Novko2018}, the presence of electron-hole pairs in the metal surface may compensate this mismatch in energy.  The IS mode can excite electron-hole pairs that experience a second scattering process with a low energy phonon mode. In the MBPT language the EMPPC is accounted by evaluating the second order intraband phonon-self-energy that reads~\cite{Novko2018,Novko2019,Bombin2023} 
\begin{widetext}
\begin{align}
    \pi_\lambda^{[2]}(\omega_{0,\lambda}; T_e;T_l)=
&-\sum_{\mu \mu^\prime\textbf{k}\sigma,\lambda^\prime\textbf{k}^\prime} \left|g^{\mu\mu}_\lambda(\textbf{k},0)\right|^2
\left(1-\frac{g_\lambda^{\mu^\prime\mu^\prime}(\textbf{k}^\prime,0)}{g_\lambda^{\mu\mu}(\textbf{k},0)}\right)
\left|g^{\mu\mu^\prime}_{\lambda^\prime}(\textbf{k},\textbf{q}^\prime)\right|^2\nonumber\\
&\times\sum_{s,s^\prime=\pm 1}
\frac{f(\epsilon_{\mu,\textbf{k}})-f(\epsilon_{\mu^\prime,\textbf{k}^\prime}-s^\prime s \omega_{\textbf{q}^\prime,\lambda^\prime})}
{\epsilon_{\mu,\textbf{k}}-(\epsilon_{\mu^\prime,\textbf{k}^\prime}-s^\prime s\omega_{\textbf{q}^\prime\lambda^\prime})}
\frac{s\left[n_b(s\omega_{\textbf{q}^\prime\lambda^\prime})+f(s^\prime\epsilon_{\mu^\prime,\textbf{k}^\prime})\right]}{\omega_{0,\lambda}\left[\omega_{0,\lambda} + i\eta +s^\prime(\epsilon_{\mu,\textbf{k}}-\epsilon_{\mu^\prime,\textbf{k}^\prime})+s\omega_{\textbf{q}^\prime\lambda^\prime}\right]},
\label{eq:secon_order}
\end{align}
\end{widetext}
where $\textbf{q}^\prime=\textbf{k}^\prime-\textbf{k}$ and $n_b(\omega_{\textbf{q},\lambda}) = 1/(e^{\beta\omega_{\textbf{q}\lambda}}-1)$ is the Bose-Einstein distribution, with $\beta=1/(k_BT_l)$ and $T_l$ the lattice temperature. Equation~\eqref{eq:secon_order} couples the studied ($\textbf{q}\approx$~0, $\lambda$) mode with other modes ($\textbf{q}^\prime$, $\lambda^\prime$) via electron-hole pairs. In the results presented in this work, the vertex correction [second term in the square bracket of Eq.~\eqref{eq:secon_order}] is neglected to reduce the computational cost. Nonetheless, it has been suggested that the contribution of the vertex correction is small~\cite{Maksimov96,Bauer1998}.

In Eqs.~\eqref{eq:first_order} and \eqref{eq:secon_order} the electronic temperature $T_e$ enters the Fermi-Dirac distribution, while the lattice temperature $T_l$ is included via the Bose-Einstein distribution that only appears in Eq.~\eqref{eq:secon_order}. This allows us to study the frequency shift and linewidth of the IS mode of CO adsorbed on the Pd(111) surface under different temperature conditions. In section~\ref{sec:thermal} we consider thermalized conditions, that is, the $T_e = T_l$ case. Instead, in section~\ref{sec:nothermal} we consider the ultrafast non-equilibrium conditions that are created when the system is excited with a pump femtosecond laser pulse. 

The theoretical treatment of the dynamical evolution of the system under the effect of an ultrafast pump pulse usually relies on a model for the degrees of freedom that are excited. Among these approaches one of the simplest and most popular techniques is the two-temperature model (TTM)~\cite{Anisimov1974,caruso22}. It  assumes that the energy of the pump pulse is initially absorbed by the electronic bath on the surface, creating hot electrons. Those hot electrons then thermalize transferring energy to other electrons or to the lattice via electron-electron and \textit{e}-ph interactions, respectively. This approximation needs as input some information about the system, namely, the electronic and lattice heat capacities, the electronic thermal conductivity, the \textit{e}-ph coupling constant and their dependence with temperature. However, the lack of experimental data for these quantities for the high electronic temperatures that are generated in pump-probe experiments forces to rely on some approximations. 
In previous works these parameters have been included in the model under some approximations, such as making use of low temperature analytical expressions and extrapolating the available experimental data to the high temperature regime. Alternatively, these quantities can be evaluated fully from first principles~\cite{Novko2019, caruso22}. The latter has been done recently by Li and Ji~\cite{Li2022} for different transition metals including Pd in a wide rage of temperatures. In this work, we make use of those results to construct a TTM. In Appendix~\ref{app:TTM} we provide further details about the TTM that we employ.

\section{The system}
\label{sec:system}
The adsorption structure of  CO on Pd(111) has stood as a puzzle for a long time. For the 0.5~ML coverage, it has been determined  that  two $c(4\times2)$-2CO arrangements coexist at temperatures in the range of hundreds of Kelvin~\cite{Rose2002,Surnev2000}: the hollow configuration CO(hollow)/Pd(111), in which the CO molecules adsorb on fcc and hcp hollow sites, and the bridge configuration CO(bridge)/Pd(111) with CO at bridge sites. 
To give insight on how much the adsorption structure of the CO adlayer may affect the vibrational dynamics, we study the two situations independently. Further details about the atomic and electronic structures for these two configurations are provided in Appendix~\ref{app:COonPd}. The electronic band structures show only small differences between the two configurations. On the contrary, clear differences arise in the vibrational spectra. For this reason, we focus on the vibrational spectra and the coupling between electronic and phononic degrees of freedom in the main text and leave the discussion about the electronic band structures for the appendices. 

\begin{figure}[tb!]
	\centering
	\includegraphics[width=1.0\linewidth]{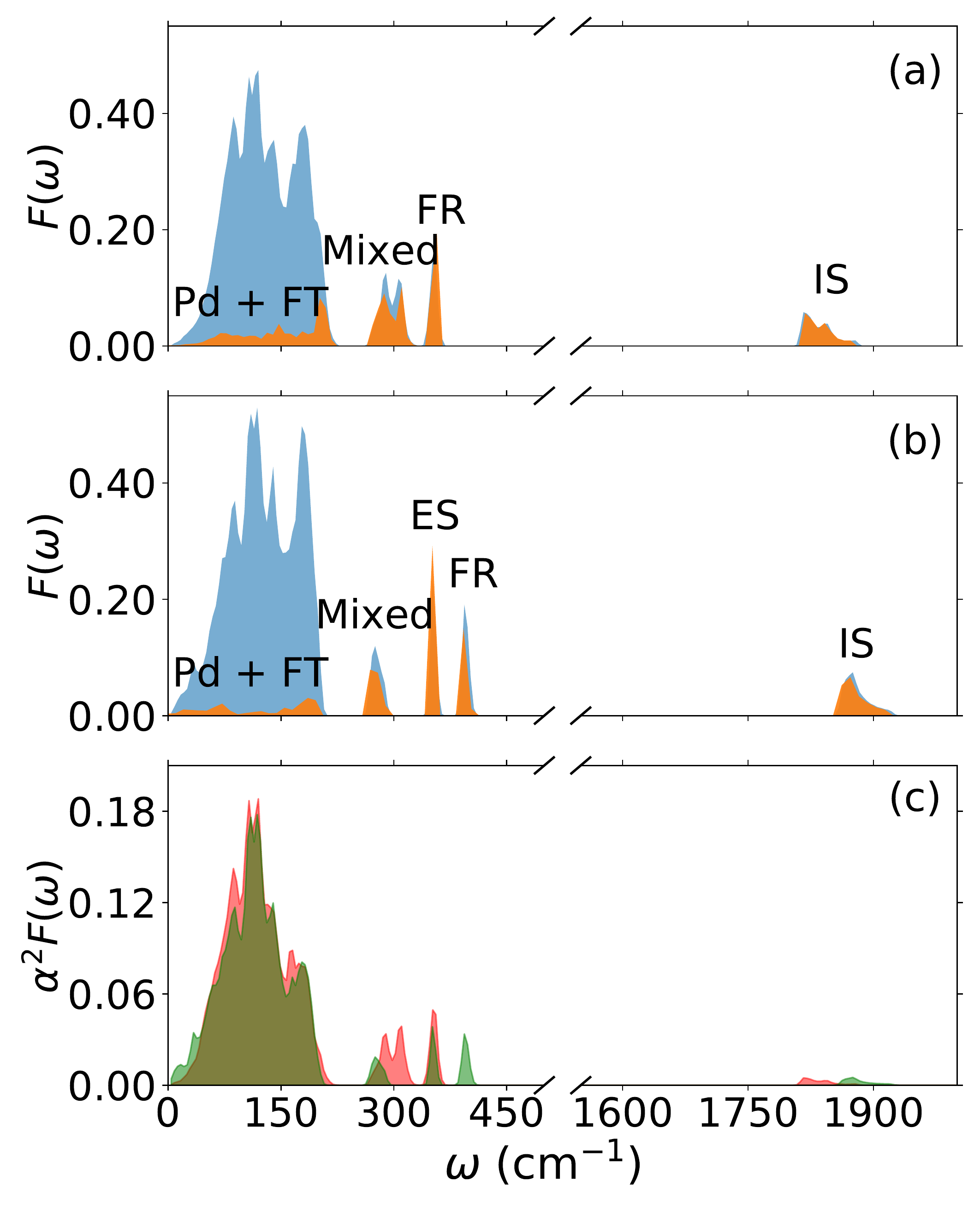}
\caption{Phonon DOS for CO/Pd(111) in (a) hollow and (b) bridge configurations. Total DOS in blue and DOS projected onto the CO modes in orange. (c) Eliashberg function for CO/Pd(111) in the hollow (red) and bridge (green) configurations. A Gaussian smearing of 0.5~meV ($\sim $~4~cm$^{-1}$) is employed.}
	\label{fig:a2f}
\end{figure}

Figures~\ref{fig:a2f}(a) and (b) show the phonon DOS for the hollow and bridge configurations, respectively. The solid blue area represents the total DOS, while the orange area corresponds to the projection onto the adsorbate normal modes. In each projected DOS, we label the energy regions in which the IS and the other vibrational modes appear, namely, external stretch (ES), frustrated rotations (FR), and frustrated translations (FT). The label ``mixed'' indicates that different adsorbate modes overlap in that energy range. The IS  mode, located at $\omega\approx1800-1900$~cm$^{-1}$, is decoupled in energy from the rest of vibrational modes (notice that the $x$-axis is shortened in the range $\omega\in[450,1600]$~cm$^{-1}$ to facilitate visualization).  The Pd modes appear in the range $\omega \in [0,230]$~cm$^{-1}$ and overlap with the CO FT modes. At intermediate energies, the vibrational spectra is dominated by FR and ES modes. It is precisely in this range where clear differences between the hollow and bridge configurations emerge. In the former case, the peak at $\omega=351$~cm$^{-1}$ corresponds to FR, while the double-peak structure around $\omega = 300$~cm$^{-1}$ is mainly a mixture of FR and ES modes. In contrast, for the bridge configuration three peaks are clearly identified in the region $\omega\in [250,415]$~cm$^{-1}$ that correspond, from the highest to lowest energy, to FR, ES, and a mixture of FT, FR, and ES modes (labeled as mixed) in the figure. 

The Eliashberg spectral function $\alpha^2 F(\omega)$, which  is shown in Fig.~\ref{fig:a2f}(c) for the hollow (red) and bridge (green) structures together, provides information on the effect of the \textit{e}-ph coupling in the phonon spectra. The Eliashberg function is here calculated as 
%
%\begin{align}
%\alpha^2F(\omega) = &\sum_{\textbf{q} \lambda}\frac{1}{\pi N(\epsilon_\mathrm{F})\omega_{\textbf{q} \lambda}}\mathrm{Im}\left[\pi_\lambda^{[1]}(\textbf{q},\omega_{\textbf{q} \lambda})\right]\delta (\omega-\omega_{\textbf{q} \lambda}) \, ,
%    \label{eqn:a2f}
%\end{align}
\begin{align}
\alpha^2F(\omega) = \frac{1}{\pi N(\epsilon_\mathrm{F})}&
\sum_{\textbf{q} \lambda}
\frac{\mathrm{Im}\left[\pi_\lambda^{[1]}(\textbf{q},\omega_{\textbf{q} \lambda})\right]}{\omega_{\textbf{q} \lambda}}\delta (\omega-\omega_{\textbf{q} \lambda}) \, ,
    \label{eqn:a2f}
\end{align}
where $N(\epsilon_\mathrm{F})$ is the DOS at the Fermi level, $\delta(x)$ is the Dirac delta function, and
\begin{equation}
\pi_\lambda^{[1]}(\textbf{q},\omega_{\textbf{q} \lambda})=\sum_{\mu \mu^\prime\textbf{k}\sigma}\left|g^{\mu\mu^\prime}_\lambda(\textbf{k},\textbf{q})\right|^2\frac{f(\epsilon_{\mu\textbf{k}})-f(\epsilon_{\mu^\prime\textbf{k} + \textbf{q}})}{\omega_{\textbf{q} \lambda} + \epsilon_{\mu\textbf{k}}-\epsilon_{\mu^\prime\textbf{k} + \textbf{q}}+ i\eta}
\end{equation}
is the $\textbf q$-dependent bare phonon self-energy for the $\omega_{\textbf{q}\lambda}$ mode. Notice that Eq.~\eqref{eq:first_order} is directly obtained by setting $\textbf{q}=0$ in the above expression.
Importantly, in order to obtain $\alpha^2F(\omega)$ one needs to calculate $\mathrm{Im}\pi_\lambda^{[1]}$ in double-delta approximation\,\cite{allen72,Novko2016,Giustino2017}.

Following common practice, the results in Fig.~\ref{fig:a2f}(c) correspond to evaluating $\alpha^2F(\omega)$ at $T=0$. 
Comparing for each structure the shape of $\alpha^2F(\omega)$ to that of $F(\omega)$ in the region where the Pd modes appear ($\omega\in[0,230]$~cm$^{-1}$) and in the CO modes region, one concludes that the coupling to the latter modes is weaker. This is inferred by noting that the relative height of the CO modes to the Pd modes in $\alpha^2F(\omega)$ is smaller than in $F(\omega)$. The coupling as defined with $\alpha^2F(\omega)$ is particularly small for the CO IS mode in which we are interested in this work, however, for our purposes this comes from the high frequency of the IS mode and not the small value of Fermi-surface-averaged phonon linewidth $\mathrm{Im}\pi_\lambda^{[1]}$. 

\section{Thermal conditions}
\label{sec:thermal}

\begin{figure}[tb!]
	\centering
	\includegraphics[width=1.0\linewidth]{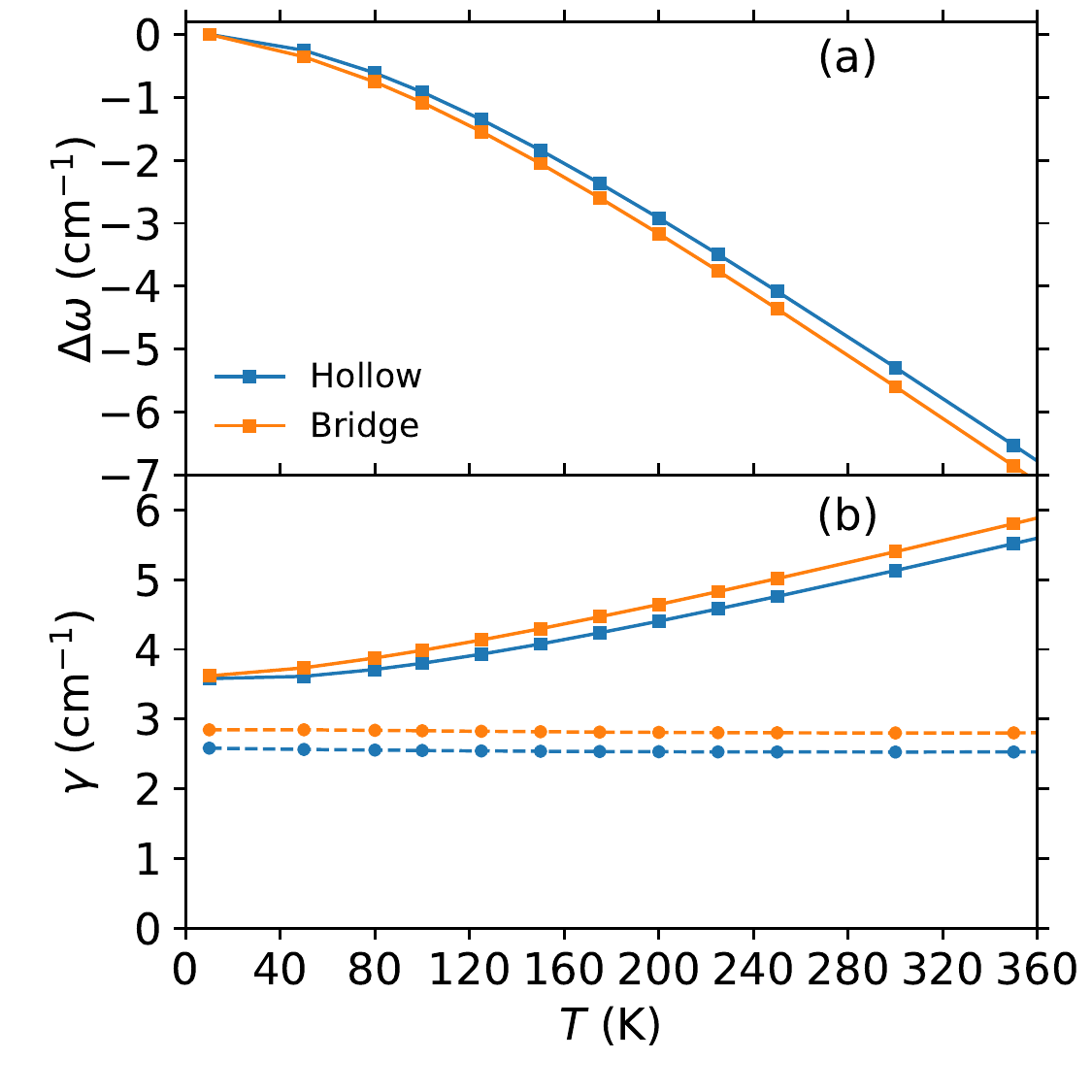}
\caption{Temperature dependence of the CO IS vibrational mode for CO(hollow)/Pd(111) (blue squares) and CO(bridge)/Pd(111) (orange squares). (a) Total frequency shift $\Delta\omega(T)$. (b) Total linewidth $\gamma(T)$. For each configuration the dashed-line connected circles show the corresponding NC contribution $\gamma^{[1]}$. 
}
	\label{fig:heat}
\end{figure}

In this section we analyze the temperature dependence of the IS mode spectra under thermal conditions ($T_e = T_l = T$). Figure~\ref{fig:heat}(a) shows the IS frequency shift $\Delta \omega(T)=\omega(T)-\omega(0)$ in the range $T\in[0,300]$~K. As expected for moderate temperatures ($k_BT \ll \omega_{0\lambda}$)~\cite{Novko2018}, the first order NC gives a negligible contribution ($\Delta\omega^{[1]}< 0.1$~cm$^{-1}$, not shown). Thus, the monotonically increasing frequency redshift that we observe in the figure as $T$ increases is driven by the EMPPC processes and, specifically, by the temperature dependence of the Bose-Einstein distribution [see Eq.~\eqref{eq:secon_order}]. 
The redshift is slightly larger for the bridge  structure, being the difference in $\Delta \omega(T)$ between both structures 0.60 cm$^{-1}$ at $T=300$~K. 

\begin{table}[!b]
\caption{\label{tab:thermal} First-order NC and second-order EMPCC contributions to the IS mode linewidth $\gamma(T)$ for the hollow and bridge configurations under two thermal conditions, $T=100$~K and 300~K. The mode-resolved contributions to EMPPC (IS dephasing, FT, ES, FR, and Pd phonons) are given in the bottom part of the table. 
All linewidths are in cm$^{-1}$.}
\begin{ruledtabular}
\begin{tabular}{ccccc}
 & \multicolumn{2}{c}{hollow} & \multicolumn{2}{c}{bridge}  \\
 \cline{2-3}\cline{4-5}
 & $T=100$~K & $T=300$~K & $T=100$~K & $T=300$~K \\
\hline
\rule{0pt}{3ex}
$\gamma^{[1]}$  & 2.55  & 2.53  & 2.83 &  2.80\\
$\gamma^{[2]}$  & 1.25  & 2.6   & 1.16 & 2.60  \\
%\hline
\rule{0pt}{3ex}
IS              & 0.025 & 0.030  & 0.021 & 0.021  \\
FT              & 0.067 & 0.130  & 0.038 & 0.082  \\
ES              & 0.23  & 0.38  & 0.17 & 0.27  \\
FR              & 0.35  & 0.54  & 0.27 & 0.40 \\
Pd              & 0.58  & 1.52  & 0.67 & 1.83 \\
\end{tabular}
\end{ruledtabular}
\end{table}
%-----------------------------------------

The total IS linewidth $\gamma$ (solid lines) and its dependence on $T$ are shown in Fig.~\ref{fig:heat}(b) for the two structures together with their corresponding first-order NC contributions $\gamma^\textrm{[1]}$ (dashed lines). The results for both structures are qualitatively very similar, the difference being the slightly larger IS phonon linewidth of the bridge structure. In each case, the NC term clearly follows the mentioned constant behavior that is expected for moderate temperatures.  This term dominates the linewidth at very low temperatures ($T<100$~K) but there is also a substantial EMPPC contribution of about 25\% that cannot be neglected. As temperature increases, the EMPPC term $\gamma^\textrm{[2]}(T)$, which is responsible for the dependence of $\gamma$ on $T$, increases with $T$ and becomes equally important. Similarly to CO/Cu(100)~\cite{Novko2018}, the dependence of $\gamma$ on $T$ goes as $T^3$ at very low temperatures and changes to linear in the range $T\in[160,300]$.
As shown in Table~\ref{tab:thermal}, the main contribution to $\gamma^\textrm{[2]}(T)$ comes from the coupling to the surface vibrational modes, followed by the coupling to the FR, ES, and FT.
Furthermore, we also find that the IS dephasing (i.e., coupling of the IS phonon at $\textbf{q} = 0$ with all other IS phonons at $\textbf{q} \neq 0$) is almost negligible.  
The relative importance of each of these contributions to $\gamma^\textrm{[2]}(T)$ is in accordance with the discussed properties of the Eliashberg spectral function. In other words, the main contributions to the EMPPC processes come from those modes that, as described by $\alpha^2F(\omega)$, couple more efficiently to electron-hole pairs. In contrast, for CO/Cu(100)~\cite{Novko2018} it was found that the coupling to the surface modes is weaker than to ES and FR modes.

Altogether, the results in Fig.~\ref{fig:heat}(b) show that the IS linewidths of the hollow and bridge structures vary, respectively, within the range $\gamma=3.55-5.1$~cm$^{-1}$ and  $\gamma=3.60-5.4$~cm$^{-1}$ as $T$ increases from 10 to 300~K. Intriguingly, very similar linewidth increase with temperature was measured for CO/Pt(111)\,\cite{Schweizer89}.
The corresponding lifetimes ($\tau = \hbar/\gamma$) at $10-150$~K for the hollow ($1.70-1.04$~ps) and bridge ($1.49-0.98$~ps) structures are comparable to the ones measured in  Pt(100) ($1.5\pm0.5$~ps~\cite{Peremans1995}), Pt(111) (2.2~ps~\cite{Beckerle1991}) and Cu(100) ($2 \pm 1$~ps~\cite{Morin1992}). The exception is the Au(111) surface where the CO is weakly physisorbed and the measured lifetime is larger ($49  \pm 3$~ps~\cite{Kumar2019,Novko2019_2}). Note also that our values of the first-order NC linewidth $\gamma^\textrm{[1]}$ are close to the theoretically obtained ones in Ref.~\cite{maurer16} for various surfaces.
Unfortunately, up to our knowledge there is no experimental data available for the CO/Pd(111) system.

\section{non-equilibrium conditions}
\label{sec:nothermal}

\begin{figure}[tb!]
	\centering
	\includegraphics[width=1.00\linewidth]{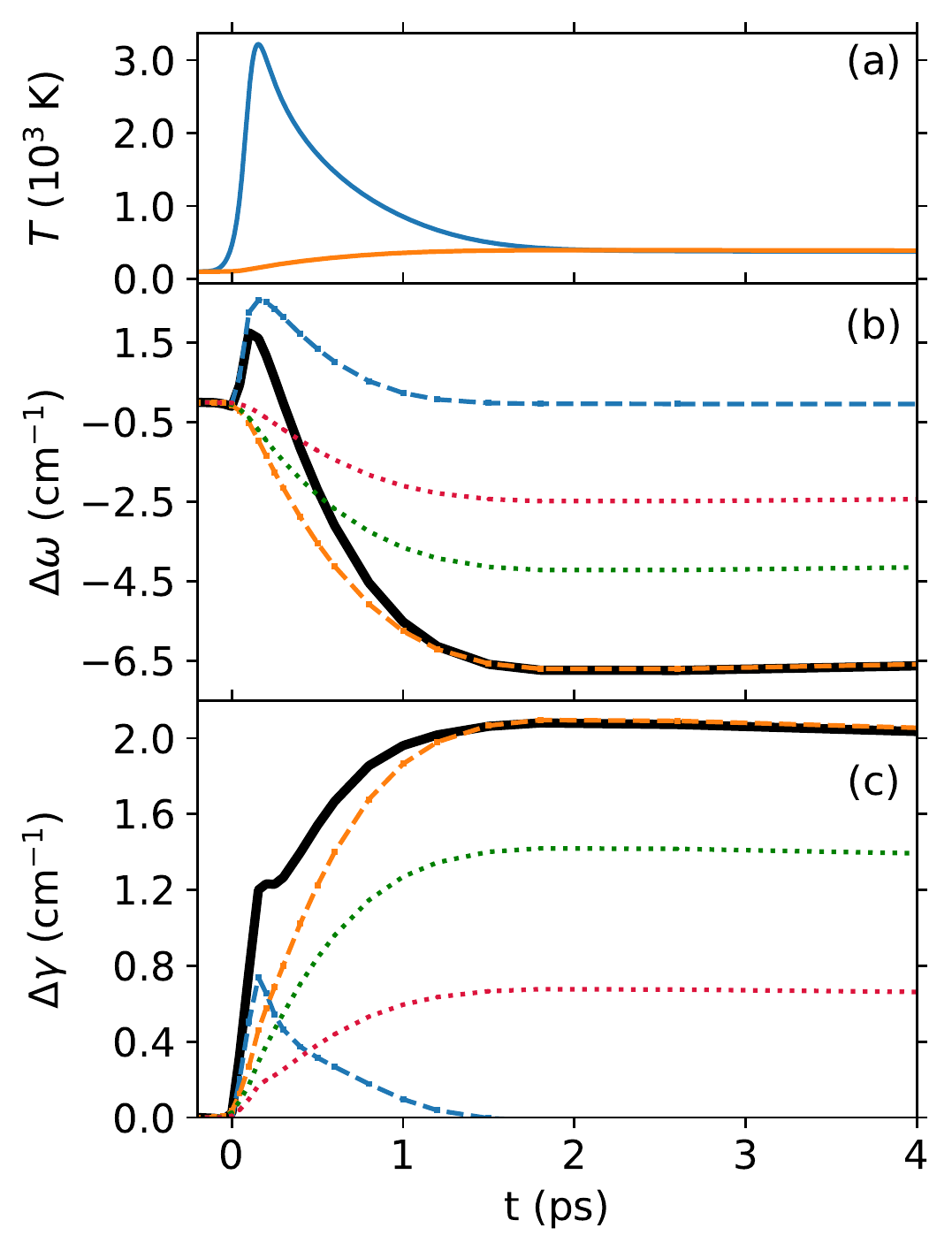}  
\caption{Transient changes in the CO IS vibrational spectra of CO(hollow)/Pd(111) induced by a 450~nm laser pump pulse with 100~fs duration, absorbed fluence 40~J/m$^2$, and peak intensity hitting the system at $t=0.1$~ps. The initial temperature is 100~K. (a) Electron $T_e(t)$ (blue) and lattice $T_l(t)$ (orange) temperatures, (b) frequency shift, and (c) linewidth change as a function of time. In (b) and (c), the black line is the sum of the first order NC term (blue dashed) plus the second order EMPPC term (orange dashed). Red and green dotted lines are the CO modes and surface modes contributions to EMPPC, respectively.}
	\label{fig:laser}
\end{figure}

In this section we assume that the CO/Pd(111) system, which is initially in thermal equilibrium at $T=100$~K, is pumped with a 450~nm laser pulse with 100~fs duration that hits the surface at $t=0.1$~ps. The considered absorbed fluences are in the range $[6,130]$~J/m$^2$ that comprises the experimental values used in existing femtosecond pump-probe vibrational spectroscopy experiments. As explained in Sec.\ref{sec:methods}, the non-equilibrium conditions created by the pump pulse are described with the TTM, from which we calculate the time-dependent electronic and phononic temperatures entering the first- and second-order phonon self-energy expressions [Eqs.\eqref{eq:first_order} and \eqref{eq:secon_order}]. Following experiments, we calculate the measured transient frequency shift due to \textit{e}-ph coupling as $\Delta\omega(t) \approx \mathrm{Re}[\pi_\lambda(\omega;t)]-\mathrm{Re}[\pi_\lambda(\omega;0)]$. Obviously, the contributions to $\Delta\omega(t)$ from the first- and second-order terms will be $\Delta\omega^{[1],[2]}(t) \approx \mathrm{Re}[\pi_\lambda^{[1],[2]}(\omega;t)]-\mathrm{Re}[\pi_\lambda^{[1],[2]}(\omega;0)$]. %
In the same manner, the transient linewidth change and corresponding first- and second-order contributions are obtained as $\Delta\gamma_{\lambda} = -2\mathrm{Im}\pi_\lambda(\omega;t)+2\mathrm{Im}\pi_\lambda(\omega;0)$ and $\Delta\gamma_{\lambda}^{[1],[2]} = -2\mathrm{Im}\pi_\lambda^{[1],[2]}(\omega;t)+2\mathrm{Im}\pi_\lambda^{[1],[2]}(\omega;0)$, respectively. 
\begin{figure*}[tb!]
	\centering
	\includegraphics[width=1.00\linewidth]{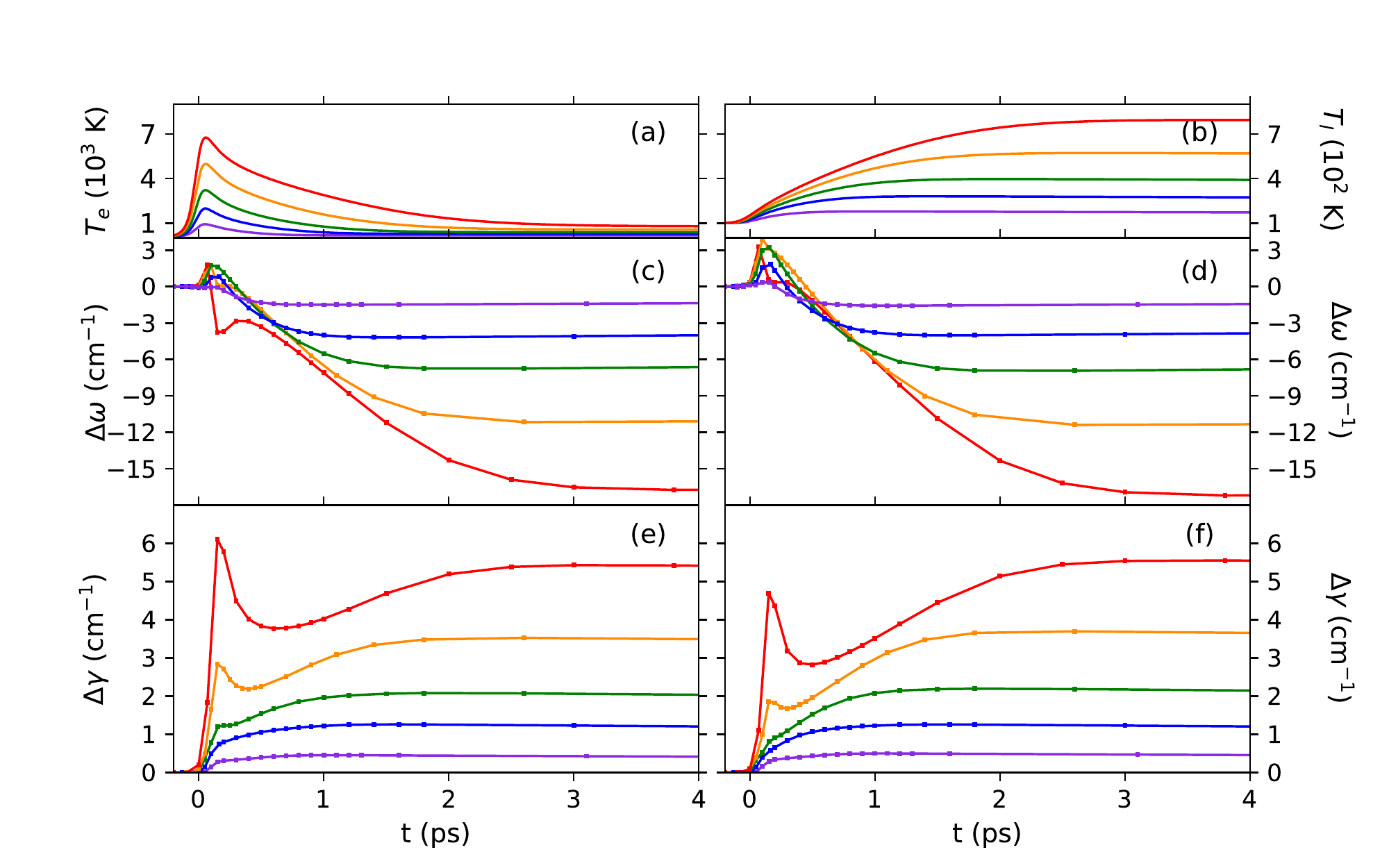}  
\caption{Transient changes in the IS vibrational spectra for CO(hollow)/Pd(111) (left) and CO(bridge)/Pd(111) (right) induced by a 450~nm laser pump pulse with 100~fs duration and peak intensity hitting the system at $t=0.1$~ps. The initial temperature is 100~K. (a) Induced electron temperature $T_e(t)$, (b) lattice temperature $T_l(t)$, (c), (d) frequency shift, and (e), (f) linewidth change as a function of time. Purple, blue, green, orange and red lines correspond to absorbed fluences of 6, 19, 40, 80, and 130~J/m$^2$, respectively.}
	\label{fig:fluences}
\end{figure*}

We start considering the case in which the pump laser deposits an absorbed fluence of 40~J/m$^2$ in the CO(hollow)/Pd(111) system. Figure~\ref{fig:laser}(a) shows the time evolution of $T_e(t)$ and $T_l(t)$. The time interval of the figure extents a few picoseconds after the instants at which $T_e$ and $T_l$ are maxima ($\approx 0.15$~ps and $\approx 2$~ps, respectively).  
Figure~\ref{fig:laser}(b) shows the transient frequency shift $\Delta\omega (t)$ that is induced in the CO IS mode (black line). It exhibits an initial blueshift that reaches a maximum value of 1.5~cm$^{-1}$ that coincides in time with the maximum electronic temperature at $t\approx 150$~fs. After the first few hundreds of femtoseconds the blueshift vanishes and gives place to a redshift that reaches its maximum at $t\approx$~1.7~ps ($\Delta\omega =-6.5$~cm$^{-1}$). The same qualitative behavior of the IS frequency shift was obtained for the bridge configuration~\cite{Bombin2023}. As shown in that case, the analysis of the $\Delta\omega^{[1]}(t)$ (blue dashed curve)  and $\Delta\omega^{[2]}(t)$ (orange dashed curve) contributions reveals that the initial blueshift is created by the first-order NC processes, while the subsequent redshift is due to the EMPPC processes. It is clear from Fig.~\ref{fig:laser}(a) that $\Delta\omega^{[1]}(t)$ and $\Delta\omega^{[2]}(t)$ follow the time evolution of $T_e$ and $T_l$, respectively. As demonstrated in Ref.~\cite{Bombin2023}, the nature of the predicted IS blueshift in CO/Pd(111) is purely electronic and related to the abrupt change in the Pd(111) electron DOS around the Fermi level. In this respect, it contrasts with the blueshift observed under thermal conditions that is explained in terms of couplings to other CO modes~\cite{Person1985,Omiya2014}. The detailed analysis that supports the electronic nature of the blueshift in the hollow structure follows the arguments of our previous work for the bridge structure~\cite{Bombin2023} and are not repeated here. 
Finally, Fig.~\ref{fig:laser}(c) shows that the transient linewidth change $\Delta \gamma(t)$ at $F=40$~J/m$^2$ starts to exhibit the two distinct regimes dictated by the different time evolution of $T_e(t)$ and $T_l(t)$. The NC term (blue dashed line) is responsible for the initial fast increase, while the slower but dominant change in the linewidth is due to the strong $T_l$ dependence of the EMPCC term (orange dashed line). Quantitatively, the largest contribution of the EMPCC processes ocurring at $t>1.5$~ps ($\sim 2.0$~cm$^{-1}$) is more than twice the maximum NC contribution of about 0.8~cm$^{-1}$ achieved at $t\sim100$~fs. 

The effect that the absorbed fluence has on the transient changes of the IS spectra for the hollow and bridge structures is shown in Fig.~\ref{fig:fluences} for the following fluences: $F = $~6, 19, 40, 80 and  130~J/m$^2$. The corresponding $T_e(t)$ and $T_l(t)$ curves are represented in panels (a) and (b), respectively. The results for the transient frequency shift of the IS mode in the hollow configuration are shown in Fig.~\ref{fig:fluences}(c). At the lowest fluence  ($F=6$~J/m$^2$), the maximum electronic temperature of about 1000~K seems insufficient to induce a non-negligible NC contribution. In this case, only a small redshift in the frequency, due to EMPP processes, is observed as the lattice temperature increases. For intermediate fluences ($F=$~19~J/m$^2$ and the above discussed 40~J/m$^2$), a blueshift appears during the first 300~fs that is progressively followed by a steady redshift of larger magnitude than the initial blueshift. Notably, for larger fluences ($F=$~80 and 130~J/m$^2$) the transient blueshift is suppressed in a faster manner than for the intermediate fluences. This behavior is a consequence of the nonmonotonic dependence of $\Delta \omega^{[1]}$ on $T_e$ that will be discussed below. Figure~\ref{fig:fluences}(d) shows the results for the transient frequency shift in the bridge structure. The time evolution of $\Delta\omega$ for the three lowest fluences ($F=$~6, 19, and 40~J/m$^2$) is qualitatively the same in the two structures, although the initial blueshift, which is due to the NC term, is quantitatively more pronounced in the bridge than in the hollow configuration. More remarkable are the differences appearing at the largest fluences, which are also due to the different $T_e$-dependence of the NC term between the two structures (see below). For $F=80$~J/m$^2$, the blueshift persists for longer times in the bridge than in the hollow configuration. For $F=130$~J/m$^2$ the blueshift is only partially suppressed in the bridge configuration, while in the hollow configuration not only the blueshift is smaller but a redshift peak appears in the time interval in which $T_e>5700$~K. 

The transient changes in the linewidth for the hollow and bridge configurations are shown in Figs.~\ref{fig:fluences}(e) and (f), respectively. Both the time evolution and the fluence dependence of the linewidth change are qualitatively the same for both structures. For the two lowest fluences, only the temperature dependence in EMPP contributes to the linewidth changes (not shown) and, as a result, $\Delta \gamma(t)$ follows the evolution of the lattice temperature. The contribution of the NC processes to $\Delta \gamma(t)$ starts to manifest at $F=$~40J/m$^2$ as an incipient peak emerging at $t\sim100$~fs that becomes more pronounced as fluence increases. In particular, for the largest fluences of 80 and 130~J/m$^2$, the well-defined peak structure that we observe for $t<0.5$~ps demonstrates that the linewidth changes in this time interval are dominated by the NC contribution. As a final remark, note that the quantitative differences between the two structures appear during the first 1-1.5~ps and are mainly due to the differences in the first order term, which is larger for the hollow structure. 

\begin{figure}[tb!]
	\centering
	\includegraphics[width=1.00\linewidth]{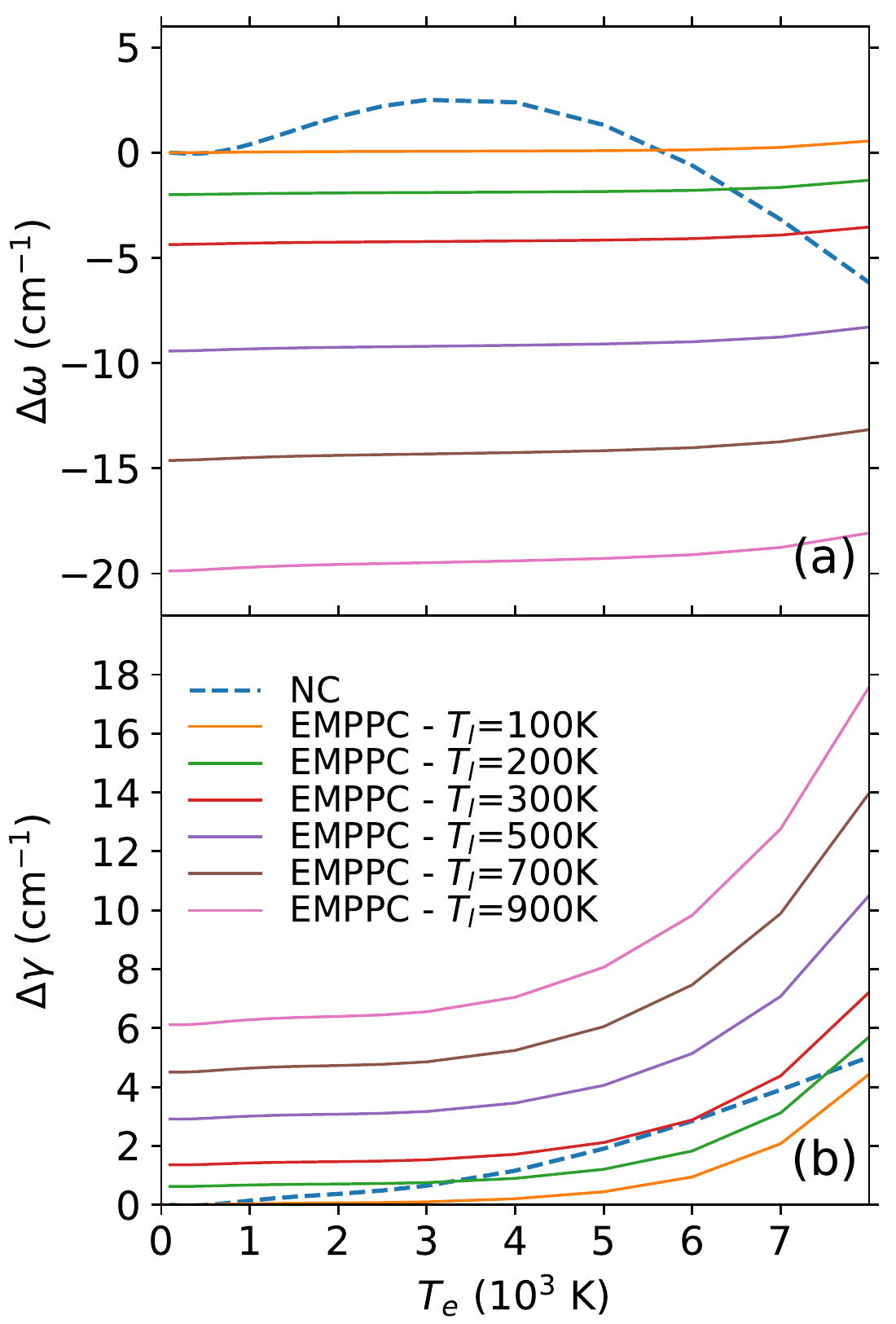}  
\caption{The NC and EMPPC contribution to the change in frequency $\Delta\omega$ and linewidth $\Delta\gamma$ of the IS mode evaluated for the CO(hollow)/Pd(111) structure as a function of $T_e$ for different $T_l$. Variations are given with respect to the thermal situation $T_l = T_e = 100$~K. Values of the EMPPC that are relevant for the regime $t<(>)1$~ps shaded in blue (red).}
	\label{fig:TeTl}
\end{figure}

To give further insight into why the electronic and lattice temperatures determine the dynamics of the IS mode in the femtosecond ($t<0.5$~ps) and picosecond ($t>1$~ps) regimes, respectively, we explicitly study the effect that $T_e$ and $T_l$ have on the NC and EMPPC terms.  Figure~\ref{fig:TeTl} shows the contributions to $\Delta\omega$ and the $\Delta\gamma$ due to the NC (dashed lines) and the EMPPC (solid lines) processes as a function of $T_e$ for the hollow configuration. In the case of the EMPPC terms ($\Delta\omega^{[2]}$ and $\Delta\gamma^{[2]}$) different lattice temperatures are considered. The ranges of temperatures shown in the figure, $T_e\in[100,8000]$~K and $T_l\in[100,900]$~K, cover the temperatures induced by the laser pulses studied in this section. 
As shown in Fig.~\ref{fig:TeTl}(a), the NC contribution $\Delta\omega^{[1]}$ produces a frequency blueshift for $T_e>900$~K that starts to decrease its  magnitude for $T_e > 3000$~K and gives place to a redshift for $T_e > 6000$~K. This is the aforementioned non-monotonic behavior that causes the rapid quenching of the initial blueshift at $F=80$ and 130~J/m$^2$ in the hollow configuration. [See in Fig.~\ref{fig:fluences}(a) that at those fluences $T_e$ goes well above 3000~K.] The dependence of $\Delta\omega^{[1]}$ on  $T_e$ is qualitatively the same for the bridge configuration but the maximum blueshift occurs at 4000~K and it turns into a redshift at about 7000~K~\cite{Bombin2023}. These qualitative differences explain that the blueshift is expected to be more pronounced and persistent in the bridge configuration [see Figs.~\ref{fig:fluences}(c) and (d)]. For fixed $T_l$, also the EMPPC term blueshifts as $T_e$ increases. This slight blueshift is however considerably smaller than the redshift of about $-20$~cm$^{-1}$ that occurs when $T_l$ increases from 100 to 900~K at fixed $T_e$. As a result, the $T_l$-induced redshift dominates the EMPPC term for any realistic pair of temperatures $(T_e, T_l)$ describing the ultrafast regime, as explicitly shown for instance by the results of Fig.~\ref{fig:laser}(b). 

Regarding the temperature dependence of the NC and EMPCC contributions to the linewidth change, Fig.~\ref{fig:TeTl}(b) shows that the NC term $\Delta\gamma^{[1]}$ is basically zero for $T_e < 1000$~K. Next, it increases monotonically with $T_e$, taking a value of $\approx4$~cm$^{-1}$ at 7000~K, which is the maximum $T_e$ reached at the highest fluence of 130~J/m$^2$ here considered [see Fig.~\ref{fig:fluences}(a)]. 
The almost constant dependence of $\Delta\gamma^{[2]}$ on $T_e$, which is expected for $k_B T_e \ll \omega_{0\lambda}$~\cite{Novko2018}, extends here up to $\sim4000$~K. Thus, for $T_e< 4000$~K, we still observe the nearly linear dependence on $T_l$ that is expected for $k_B T_e \ll \omega_{0\lambda}$ (see Sec.~\ref{sec:thermal}). Actually, we observe this linear dependence for the whole lattice temperature range shown in the figure, even if the dependence of the EMPPC term on $T_e$ is visibly enhanced for $T_e > 4000$~K. That enhancement is due mainly to the increasing importance that the IS dephasing process acquires at such high $T_e$. 
In this respect, it is worth noticing that the peaks observed in $\Delta \gamma$ at $t=0.1$~ps for the largest fluences of 80 and 130~J/m$^2$ [see Fig.~\ref{fig:fluences}(e)] are not only caused by the NC term, as it is the case for $F= 40$~J/m$^2$ [see Fig.~\ref{fig:laser}(c)], but also by the EMPPC term via the IS dephasing.

\section{Comparison to other CO/metal systems} 
\label{sec:comparison}

The thermal~\cite{Novko2018}  and non-equilibrium~\cite{Novko2019} IS vibrational spectra of CO adsorbed on the Cu(100) surface have been studied within the same theoretical framework that we use here. But before comparing the results of the IS vibrational spectra between the two systems, it is worth analyzing the intrinsic differences between the Cu and Pd surfaces that are likely to affect the adsorbate vibrational spectra. First, the laser-induced $T_e$ ($T_l$) for equal absorbed fluences is lower (higher) in Pd(111) than in Cu(100) because of the larger \textit{e}-ph coupling constant in bulk Pd ($\lambda=$~0.4) than in bulk Cu ($\lambda=$~0.14). 
Second, the electron DOS around the Fermi level  are very different for the two surfaces. Whereas in Cu(100) the $sp$-band crosses the Fermi level and the DOS above and below $\varepsilon_F$ is rather constant, in Pd(111) the Fermi level lies at the edge of the $d$-band continuum and, as a result, the DOS is much larger below than above $\varepsilon_F$ (see~Fig.~\ref{fig:bandsFCCHCP}). As already discussed, the latter has important implications for the laser-induced IS dynamics due to the significant change of the chemical potential with temperature~\cite{Bombin2023}. 

Under thermal conditions, the IS linewidth at $T=10$~K for CO/Cu(100) ($\gamma\approx 2.9$~cm$^{-1}$~\cite{Novko2018}) is slightly smaller than for CO/Pd(111) ($\gamma\approx3.6$~cm$^{-1}$). 
Interestingly, the first-order contribution $\gamma^{[1]}$ is larger in CO/Pd(111) probably due to the larger DOS around $\varepsilon_F$. In contrast, the second-order contribution $\gamma^{[2]}$ is smaller due to a weaker electron-mediated coupling of the IS mode to the Pd(111) phonons. A possible explanation is the largest mass of the Pd atoms with respect to the Cu mass (notice that $g_\lambda^{\mu,\mu^\prime}\sim 1/\sqrt{M_\lambda}$ with $M_\lambda$ the mass associated with the phonon $\lambda$). Finally, the change of the IS linewidth with temperature in Cu(100) and Pd(111), namely, $\Delta \gamma (T)\approx 0.9$~cm$^{-1}$ and 0.8~cm$^{-1}$, respectively, for $T\in[10,150]$~K,  are very similar. This temperature dependence is dominated in both systems by the coupling with the surface modes. We note once again that the measured linewidth increase of about 3\,cm$^{-1}$ between 50\,K and 300\,K for CO/Pt(111)\,\cite{Schweizer89} is close to our results for CO/Pd(111).

Regarding the properties of the transient vibrational spectra induced by femtosecond laser pulses, there are significant differences between CO/Cu(100) and our results for CO/Pd(111). One important difference, which is directly related to the small $T_l$ that the laser-excited electrons induce in Cu(100) as compared to Pd(111), refers to the time evolution of the transient spectra. The largest frequency and  linewidth changes in CO/Cu(100) occur during the initial $T_e$-driven subpicosecond regime in which $T_e\gg T_l$. In the subsequent $T_l$-driven picosecond regime (i.e., when $T_e\simeq T_l$), the temperatures are typically much smaller (e.g., about 400~K for $F=170$~J/m$^2$) and so the frequency and linewidth changes. In contrast, the spectral changes in CO/Pd(111) are more pronounced during the steady picosecond regime that in this system is characterized by large lattice temperatures (e.g., $\sim$1000~K for $F=130$~J/m$^2$). Only for the largest fluences (with $T_e > 4000$~K), the transient changes in the fast and steady regimes are comparable. Furthermore, there are additional qualitative differences between the two systems that it is worth remarking. In CO/Cu(100)~\cite{Novko2019}, the pronounced IS frequency redshift that occurs during the first picosecond is ruled by the first-order NC processes. 
The transient linewidth change however is dominated by the EMPPC processes. In particular, the largest change is due to the IS dephasing that takes place within the first picosecond. 
In CO/Pd(111), an initial frequency blueshift driven by the NC processes is expected to occur during the first 0.3~ps for absorbed fluences large enough to raise $T_e$ above 1000~K. The blueshift progressively turns into a large and steady redshift, which is caused by the EMPPC processes leading the late picosecond regime. The latter dominates the linewidth change. It is only for the largest fluences ($T_e>4000$~K) that the subpicosecond regime controlled by the NC processes is also visible. 

It is also meaningful to compare our results to time-resolved vibrational spectra measured in other surfaces with pump-probe vibrational spectroscopy. 
In Ref.~\cite{Watanabe2010}, the experimental linewidth change of the IS mode of CO on Pt(111) exhibits the same behavior that we obtain here for CO/Pd(111). Specifically, the $T_l$-driven late steady regime appears at all the experimental fluences (40, 80, and 130~J/m$^2$), while the fast $T_e$-driven subpicosecond regime becomes increasingly important at $F \gtrsim 80$~J/m$^2$. Regarding the frequency change, the late steady frequency redshift and its dependence on $F$ are also quite similar to our CO/Pd(111) results. However, a redshift instead of a blueshift seems to occur in the subpicosecond regime. Nonetheless, there is an unusual nonmonotonic dependence of $\Delta\omega$ on the absorbed fluence as found in later experiments with improved subpicosecond resolution~\cite{Inoue2012}. We speculate that the differences between CO/Pt(111) and CO/Pd(111) may be caused by the differences in their electron DOS. Even if qualitatively the DOS of the two systems are similar, the decay of the $d$-band edge above $\varepsilon_F$ is sharper in Pd(111) than in Pt(111). Thus, we may expect a somehow softer dependence of the chemical potential on temperature that would affect the frequency sift behavior. 

The fact that the transient IS mode spectra follows the dynamics of the lattice temperature in the picosecond regime is common to other experiments in transition metal surfaces. Lane \textit{et al.} experimentally studied the ultrafast vibrational dynamics  of CO and NO on the Ir(111) surface~\cite{Lane2006,Lane2007}. The authors found that the IS frequency shift of NO (CO) follows the time evolution of the lattice temperature when the absorbed fluence is below 8~J/m$^2$ (20~J/m$^2$). Similarly, in the case of CO/Ru(1000) it has been shown that the IS dynamics follows the lattice temperature at least up to a fluence of 55~J/m$^2$~\cite{Bonn2000}.

Finally, it is worth remarking the blueshift observed for CO molecules coordinated to ruthenium tetraphenylporphyrin adsorbed on Cu(110) with pump-probe vibrational spectroscopy~\cite{Omiya2019}. The values of the blueshift at different fluences are very similar to the results presented here for CO/Pd(111).  For sufficiently small fluences, only a redshift that evolves in time with the lattice temperature is observed. However, for larger values of the fluence, a blueshift in the subpicosecond regime emerges.

\section{Conclusions}
\label{sec:conclusions} 

In summary, we have studied the vibrational relaxation of the internal stretch mode of CO adsorbed on the Pd(111) surface using density functional theory in combination with many-body perturbation theory. In particular, by evaluating the phonon self-energy up to second order in the electron-phonon (\emph{e}-ph) coupling matrix elements, we have characterized the two mechanisms that participate in the vibrational relaxation of the CO internal stretch mode: the first-order interband nonadiabatic coupling (NC) and the second-order intraband electron mediated phonon-phonon (EMPPC) coupling. Under thermal conditions both mechanisms contribute significantly to the linewidth $\gamma(T=0)\approx 3.6$~cm$^{-1}$, with the interband contribution $\gamma^{[1]} (T=0)\approx 2.7$~cm$^{-1}$ being the largest. The redshift and change in the linewidth that are induced as temperature increases are explained mainly in terms of the coupling of the IS mode to the low energy phonon modes of the system. We have also studied the subpicosecond and picosecond transient changes that are induced in the IS phonon frequency and linewidth when the CO/Pd(111) system is excited with a femtosecond laser pulse (non-equilibrium conditions). 
For large electronic temperatures (with $T_e\gg T_l$), the particular electronic structure of Pd may give rise to an effective screening of the \textit{e}-ph interaction, inducing an unconventional blueshift in the IS frequency. Thus, an initial rapid blueshift lasting a few hundreds of femtoseconds, followed by a larger and steady redshift is predicted to occur for absorbed fluences $F\gtrsim 19$~J/m$^{2}$. The transient blueshift is exclusively driven by the coupling to hot carriers, while the subsequent redshift is due to coupling to other phonons via the EMPPC mechanism. In what concerns to the transient increase of the linewidth, the NC processes are responsible for the fast initial increase that can be observed in the initial few hundreds of femtosecond at sufficiently large fluences for which $T_e$ rises above 1000~K. In contrast, the EMPPC processes govern the subsequent steady regime that follows the time evolution of $T_l$. 
Finally, we have shown that the transient IS vibrational spectra of the hollow and bridge adsorption structures that have been observed at 0.5~ML of CO coverage are qualitatively rather similar. It is only for the largest fluences that the behavior of the frequency shift occurring during the initial 0.5~ps differs. 

\appendix
% Here se describe the Pd slab
\section{Palladium slab}
\label{app:Pdslab}
\begin{figure}[tb!]
	\includegraphics[width=1.00\linewidth]{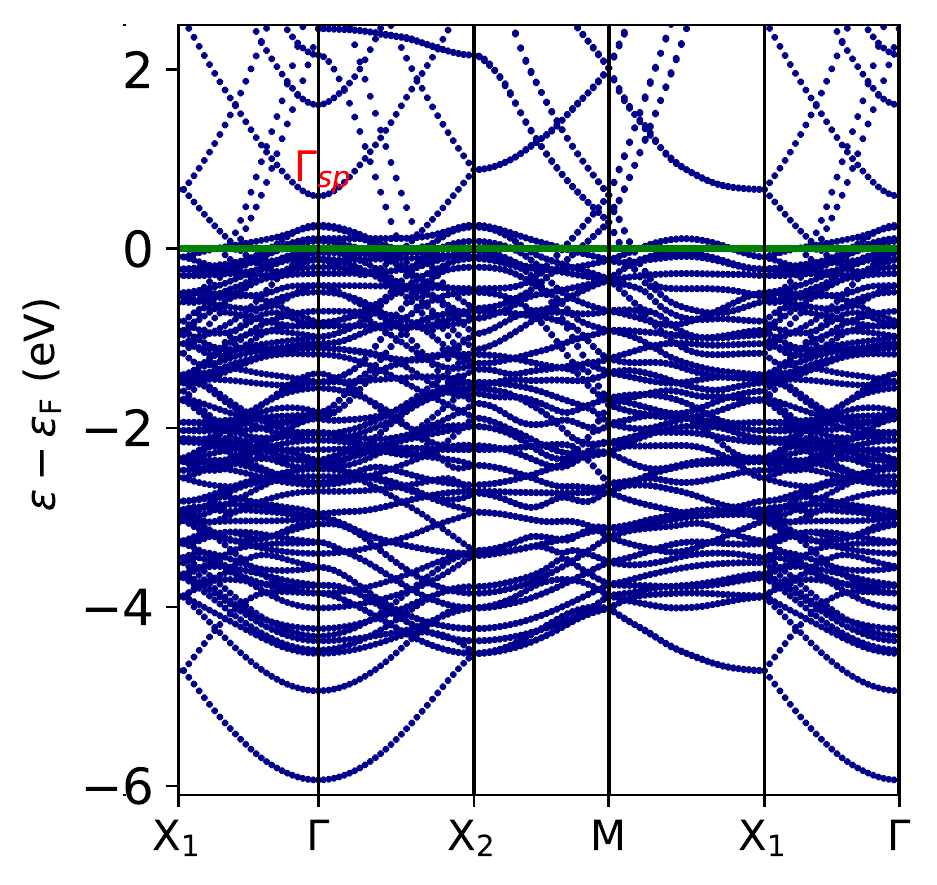}
\caption{Band structure of the Pd(111) surface calculated in the $c(2\times \sqrt{3})$rect supercell.}
	\label{fig:bands_surface}
\end{figure}

The pristine Pd(111) surface is modeled using a periodic four-layer slab and the surface supercell depicted in Fig.~\ref{fig:adsorp_sites}. We include 12.9~{\AA} of vacuum to ensure negligible interaction between the slab and its periodic images. This supercell is the minimum cell that allows us to study the two coexisting $c(4\times2)$-2CO adsorption structures that are reported in experiments for a CO coverage of 0.5~ML (see Appendix~\ref{app:COonPd} for details). The relaxed structure is obtained by minimizing all the atomic forces below 2$\times$10$^{-5}~$~Ry/$a_\mathrm{B}$. After relaxation the optimized lattice constant is $a=3.969$~{\AA} in close agreement with the experimental value $a=3.89$~{\AA}~\cite{haynes2016crc}. The interlayer distance between the two inner layers (2.30~{\AA}) is close to that of bulk Pd ($a/\sqrt{3}\sim 2.292 $~{\AA}). However, it is known that this quantity has a slow, nonmonotonic convergence with the number of layers (see, for example, Ref~\cite{Benedek2010}). Due to the computational cost of the calculations that we present in this work, we restrict the number of Pd layers to four.

The band structure of the Pd slab is shown in Fig.~\ref{fig:bands_surface}. It has been interpolated using 96 maximally localized Wannier functions. Previously, the electronic states and charge density employed to obtain the Wannier functions were computed self-consistently on a 12$\times$12$\times$1 $\Gamma$-center Brillouin zone grid. Due to the supercell that we are using, the $\Gamma$-X$_1$=(1,0,0) and $\Gamma$-X$_2$=(0,1,0) directions are nonequivalent, as in real space they correspond to the  directions connecting first and second neighbors on the Pd(111) surface. In other words, the $\Gamma$-X$_1$ direction corresponds to a high-symmetry direction in the irreducible Brillouin zone of the Pd(111) surface, but the $\Gamma$-X$_2$ does not. In the plot we label the lowest energy parabolic-like excitations that appear around the $\Gamma$ point at about 1~eV above the Fermi level as $\Gamma_{sp}$. As shown in Appendix~\ref{app:COonPd}, those states will hybridize with the unoccupied CO orbitals upon CO adsorption on Pd(111).  

% CO adsorption structures
\section{CO Adsorption structures}
\label{app:COonPd}
\begin{table}[b]
\caption{CO properties of the optimized $c(4\times2)$-2CO structures corresponding to 0.5~ML of CO adsorbed on Pd(111):  Adsorption energy per molecule $E_a$, bond length d(C-O), adsorption height of the CO geometrical center $Z$ (respect to the Pd topmost layer), polar angle of the molecular axis respect to the surface normal $\vartheta$, and IS ($\textbf{q}=0$) phonon frequency $\omega_{0}$.}
\label{table:CO_adsor}
\begin{ruledtabular}
\begin{tabular}{ l c c}
	& hollow & bridge\\  \cline{2-3}
	$E_a$ (eV)	& -1.453& -1.409 \\
	d(C-O) (\AA) &1.169, 1.169 & 1.164 \\
	$Z$ (\AA)	& 1.977, 1.987& 2.046 \\
	$\vartheta$  &  2.89$^\circ$,2.73$^\circ$ & 0.53$^\circ$ \\
	$\omega_{0}$ (cm$^{-1}$) & 1883 & 1924\\
\end{tabular}
\end{ruledtabular}
\end{table} 

\begin{figure}[tb!]
	\centering
	\includegraphics[width=1.00\linewidth]{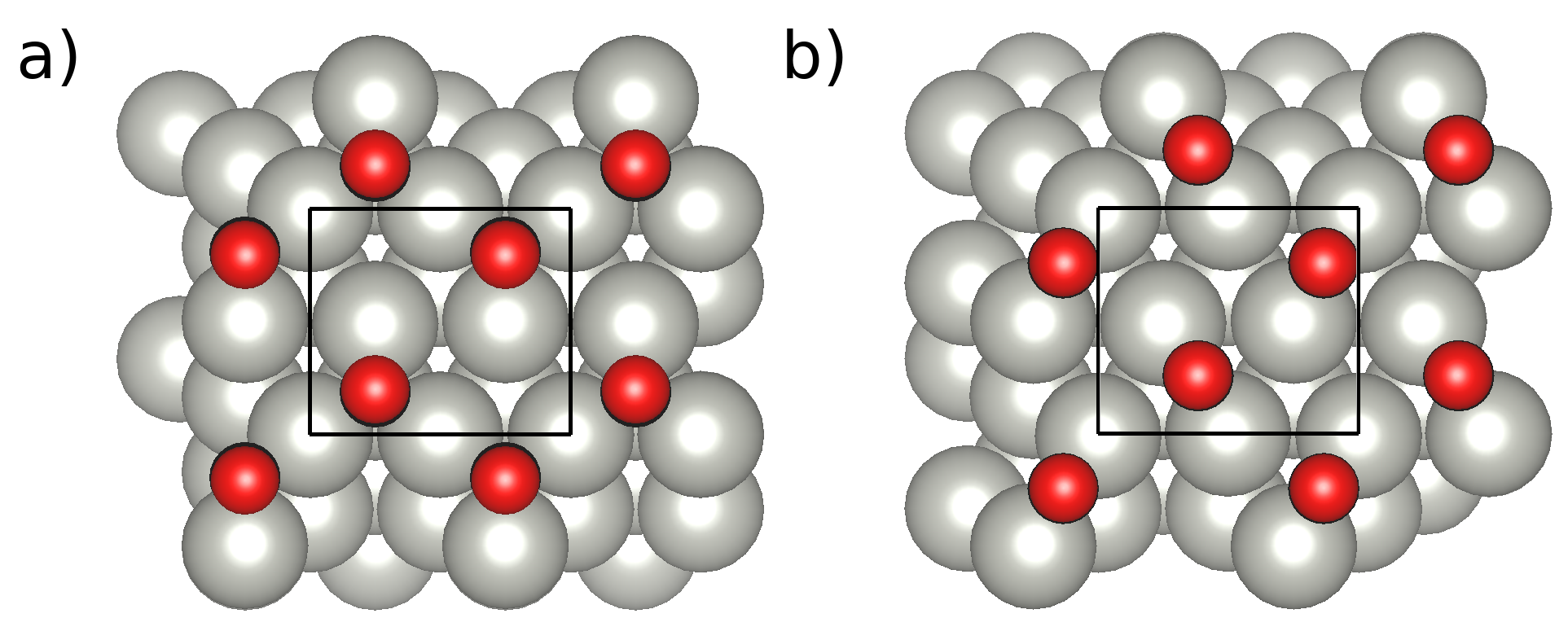}
\caption{Top view of the Pd(111)-$c(4\times2)$-2CO structures: (a) hollow configuration with CO at fcc and hcp sites and (b) bridge configuration with CO at bridge sites. Pd, C, and O atoms colored in gray, black, and red, respectively. The cell used in the calculations is indicated with solid black lines. }
	\label{fig:adsorp_sites}
\end{figure}

The CO$_{\mathrm{(gas)}}$ bond length calculated with the BEEF-vdW functional is 1.135~{\AA}. It  reproduces the experimental value of 1.128~{\AA}~\cite{NIST} within a relative error of 1\%. The calculated  frequency of 2154.6~cm$^{-1}$ for CO$_{\mathrm{(gas)}}$ also agrees well with the experimental value of 2143.16~cm$^{-1}$~\cite{Nakamoto2006}.

The precise CO adsorption sites in the stable $c(4\times2)$ structure at 0.5~ML coverage on Pd(111) has been the subject of a long debate~\cite{Vasiliy2003,Rupprechter2007}. Early infrared reflection-absorption spectroscopy studies assigned the bridge adsorption site~\cite{Bradshaw78}. Gie{\ss}el \textit{et al.} performed photoelectron diffraction studies and concluded that CO adsorbs in a mixture of fcc and hcp sites~\cite{Giebel1998}.  According to subsequent X-ray photoelectron spectroscopy experiments, the fcc+hcp structure prevail at 120~K, while the two proposed structures (fcc+hcp and bridge) coexist at 300~K~\cite{Surnev2000}. Nonetheless, scanning tunneling microscopy images showed later on that the hollow and bridge structures can also coexist at 120~K~\cite{Rose2002}. 

In view of this debate, we started our theoretical study by analyzing the stability of different Pd(111)-$c(4\times2)$-2CO structures that differ in the CO adsorption sites. Our DFT+BEEF-vdW calculations predict the following stable structures: fcc+hcp, 2fcc, fcc+bridge, hcp+bridge and 2bridge, with $-$1.453, $-$1.376, $-$1.391 and $-$1.409\,eV being,  respectively, the corresponding adsorption energy per molecule. In contrast, the 2hcp configuration is not stable, since it ends up into the fcc+hcp structure upon relaxation. Previous DFT studies also predict  that fcc+hcp is the most stable structure at low temperature (see for example Ref.~\cite{Hooshmand2017} for a study with the optB88-vdW functional). Our results predict that the bridge structure is only 44~meV more energetic than fcc+hcp. For this reason, and following the experimental evidence, in this work we study both structures. 
The top view of the two relaxed structures is shown in Fig.~\ref{fig:adsorp_sites}.

Table~\ref{table:CO_adsor} summarizes the adsorption energy per molecule $E_a$, the CO bond length d(C-O),  the height $Z$ of the CO geometrical center from the surface (defined as the average heights of the Pd atoms in the topmost layer), the polar angle of the CO molecular axis with respect to the surface normal $\vartheta$, and the CO IS phonon frequency $\omega_0$ at the adsorption position for the two structures. For the hollow structure the CO located at the fcc (hcp) site is adsorbed at a distance  $Z =$~1.977 (1.987)~{\AA} with a tilting angle $\vartheta =$~2.89$^\circ$ (2.73$^\circ$). These results are similar to those reported in previous DFT studies\,\cite{Hooshmand2017}. The structural parameters of the bridge configuration do not differ much from those of the hollow one. The most apparent change is the reduction of the tilting angle $\vartheta$ from $\sim 3^{\circ}$ to 0.53$^\circ$. 
The main difference between the two structures arises in the value of the CO IS frequency, being 1883 and 1924~cm$^{-1}$ for the hollow  and bridge structures, respectively. These values have to be compared to the existing experimental frequencies %measured for the 0.5~ML coverage 
1920~cm$^{-1}$ (200~K)~\cite{Giebel1998} and 1936~cm$^{-1}$ (300~K)~\cite{Bradshaw78}.

\begin{figure}[tb!]
	\centering
	\includegraphics[width=1.0\linewidth]{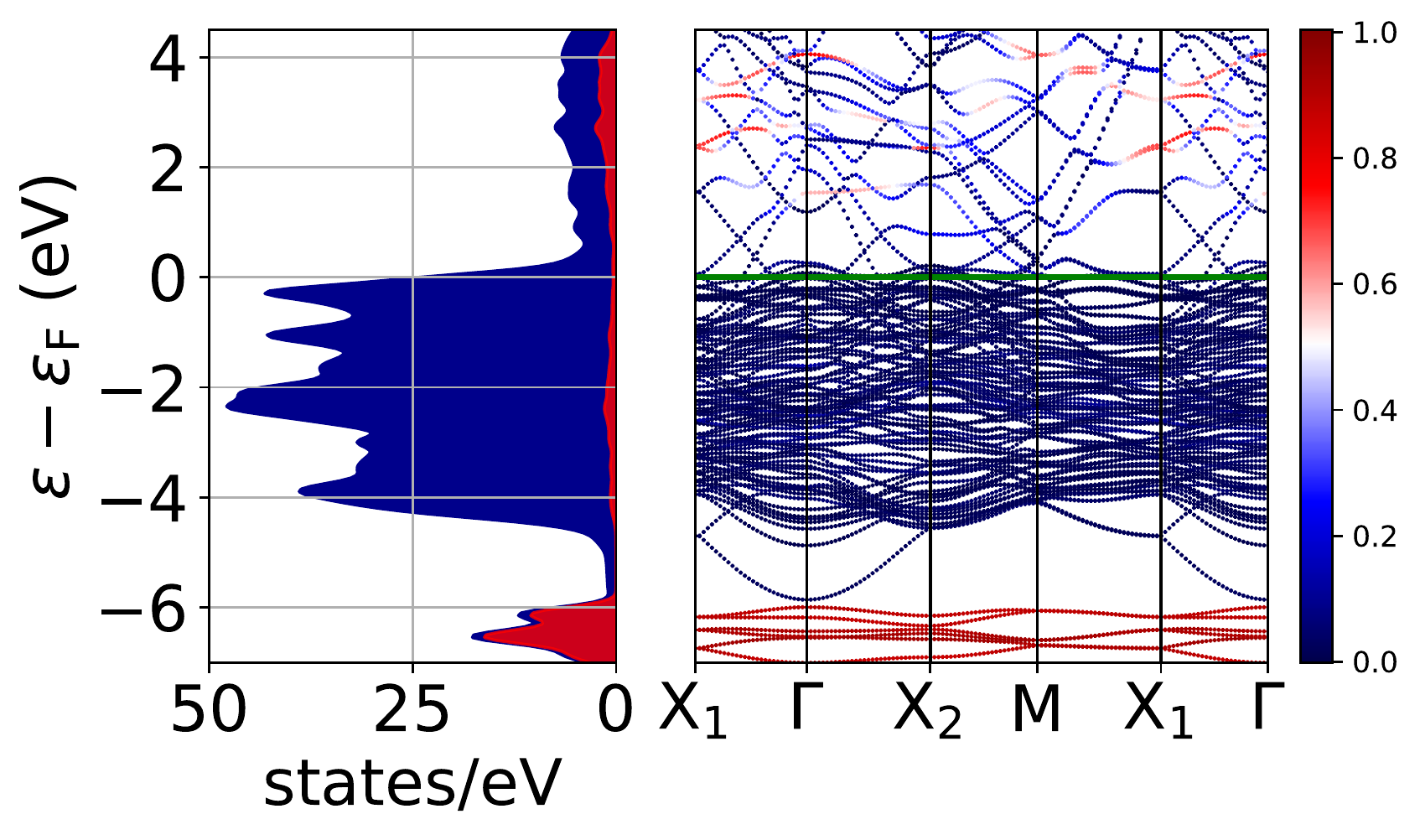}
\caption{Pd(111)-$c(4\times2)$-2CO$_\textbf{fcc+hcp}$ DOS (left) and band structure (right). Total DOS in blue and the projection onto the CO localized Wannier functions in red. In the band structure the Fermi level is in green and the weight of the CO-Wannier states is encoded in the color bar.} 
	\label{fig:bandsFCCHCP}
\end{figure}

\begin{figure}[tb!]
	\centering
	\includegraphics[width=1.0\linewidth]{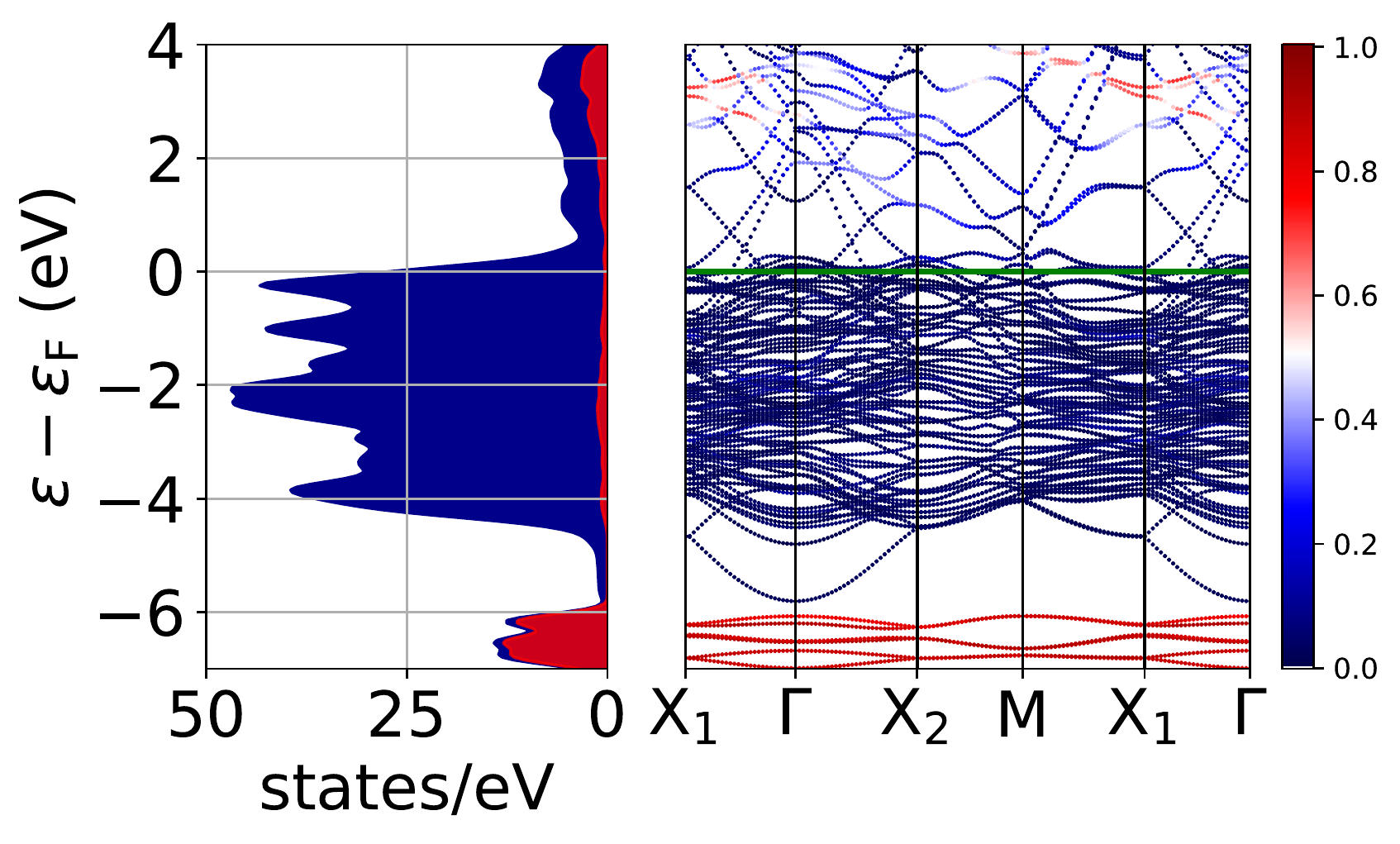}
\caption{Same as Fig.~\ref{fig:bandsFCCHCP} for Pd(111)-$c(4\times2)$-2CO$_\textbf{bridge}$.}
	\label{fig:bandsbridge}
\end{figure}

The CO/Pd(111) band structure and electron DOS for the  hollow and bridge structures are shown in Figs.~\ref{fig:bandsFCCHCP} and \ref{fig:bandsbridge}, respectively. Both the DOS and the band structure are projected onto the Wannier functions that are localized around the CO molecules. For simplicity we refer to them as CO Wannier states. In both structures the main contribution of the unoccupied CO states starts at about 2~eV above $\varepsilon_{\mathrm{F}}$. By comparing the Pd(111) band structure to those of CO/Pd(111), we observe that the $\Gamma_{sp}$ states of the Pd surface are shifted to a higher energy after CO adsorption. Strong hybridization of the folded  parts of these bands with the CO empty states is observed around the $\mathrm{X}_1$ and $\mathrm{X}_2$ high-symmetry points. This hybridization is stronger in the hollow than in the bridge structure.

\section{Two temperature model}\label{app:TTM}
The two temperature model~\cite{Anisimov1974} is widely used to describe the interaction of intense short laser pulses with metal surfaces~\cite{frischkorncr06,saalfrank2006,caruso22}. This interaction is described in terms of the dynamics of the energy exchange between the laser and the metal surface. The latter, is modeled by two coupled thermal baths: one standing for the electrons, characterized by temperature $T_\mathrm{e}(t)$; and the other standing for the lattice atoms, characterized by temperature $T_\mathrm{l}(t)$. 
Note that under realistic experimental conditions, the laser initially creates non-equilibrium electron and phonon distributions. 
It has been shown that for semiconductor systems, a small signature of these initial non-thermal distributions can persist for relatively long times~\cite{Karlsson2021}. 
In the present case, considering the small lifetime of electrons in Pd (around $10$~fs~\cite{Zhukov2002}), the thermalization is expected to occur fast and these effects are expected to be small.
The interbath coupling is assumed to depend linearly on the difference between both temperatures, i.e., $G (T_\mathrm{l}-T_\mathrm{e})$, with $G$ being the \textit{e}-ph energy exchange coupling constant of the metal. Also, as stated in the main text, the energy of the laser pulse is assumed to be entirely absorbed by the electron bath. With further standard assumptions such as considering that the incoming laser fluence is constant in the plane parallel to the metal surface,  and neglecting lattice thermal diffusion into the bulk, the resulting canonical TTM heat transfer equation reads
\begin{equation}
\begin{split}\label{eq:TTM}
    C_\mathrm{e}\frac{\partial T_\mathrm{e}}{\partial t} &=
        \frac{\partial}{\partial z}\left(\kappa_\mathrm{e}\frac{\partial T_\mathrm{e}}{\partial z}\right)-G(T_\mathrm{e}-T_\mathrm{l})+S\\
    C_\mathrm{l}\frac{\partial T_\mathrm{l}}{\partial t} &=
        G(T_\mathrm{e}-T_\mathrm{l}),
\end{split}
\end{equation}
where $C_\mathrm{e}$ and $C_\mathrm{l}$ are the electron and lattice heat capacities, respectively; $\kappa_\mathrm{e}$ is the thermal conductivity of the electron subsystem used in its simplified form~\cite{Ashcroft1988} $\kappa_\mathrm{e}(T_\mathrm{e},T_\mathrm{l})=\kappa_0\frac{T_\mathrm{e}}{T_\mathrm{l}}$ (which is a reasonable approximation for  $\kappa_\mathrm{e}$  for $d$-metals, such as Ni and Fe~\cite{Petrov2013}) with $\kappa_0$ being the room temperature (300~K) thermal conductivity of palladium; $z$ denotes the spatial direction perpendicular to the surface plane; and $S$ is the laser source term here modeled as
\begin{equation}
    S(z,t)=\frac{\alpha_\lambda F}{2\Delta t}\mathrm{e}^{-\alpha_\lambda z}
    \mathrm{sech}^2\left(\frac{t}{\Delta t}\right),
\end{equation}
with $\alpha_\lambda$ being the light absorption coefficient of the metal for a source of wavelength $\lambda$, $\Delta t=0.5\sigma_\mathrm{FWHM}/\mathrm{arccosh}{\sqrt{2}}$ the full width at half maximum of the temporal profile of the laser pulse ($\sigma_\mathrm{FWHM}=100$~fs) multiplied by a constant factor, and $F$ the absorbed fluence. The absorption coefficient $\alpha_\lambda$ is calculated as $\alpha_\lambda=\frac{4\pi k_\lambda}{\lambda}$, with $k_\lambda$ being the extinction coefficient of the material at wavelength $\lambda$. In order to describe temperature dependence of $C_\mathrm{l}$, we have used the same expression as in a previous TTM laser-induced desorption study of O$_2$ from a Pd(111) surface~\cite{Szymanski2007}, i.e.,
\begin{equation}
    C_\mathrm{l}(T_\mathrm{l}) = 
    C_\mathrm{2}+\frac{C_\mathrm{1}-C_\mathrm{2}}{1+(T_\mathrm{l}/T_\mathrm{0})^p}+mT_\mathrm{l},
\end{equation}
being $C_\mathrm{1}$, $C_\mathrm{2}$, $T_\mathrm{0}$, $p$, and $m$ fitting coefficients optimized to reproduce measurements compiled in Ref.~\cite{Miiller1971} for temperatures below 1000~K (Fig. 7 therein). All the constants and fitting values used in our model are summarized in Table~\ref{tab:TTM}.
\begin{figure}[tb!]
	\centering
	\includegraphics[width=1.0\linewidth]{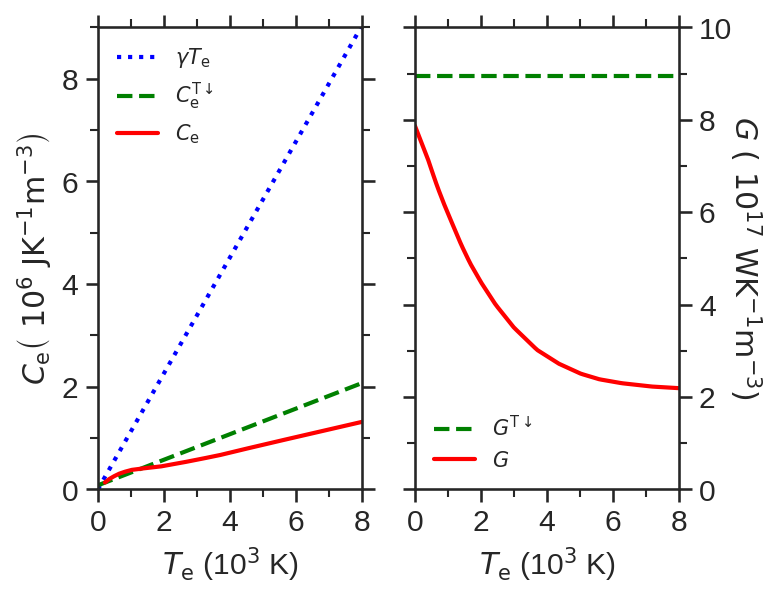}
    \caption{Electronic temperature dependence of the electron heat capacity (left panel) and the \textit{e}-ph energy exchange rate (right panel) of palladium according to different models present in the literature. In red, we show the DFT-based models we have used in this work, extracted from Ref.~\cite{Li2022}. In blue and green, we show models adjusted to experimental data for temperatures below 50~K and  2000~K, respectively, extracted from Ref.~\cite{Szymanski2007}}
	\label{fig:CeG}
\end{figure}

\begin{table}[h!]
\begin{ruledtabular}%\raggedright
    \caption{Constants and fitting coefficients used in the TTM model}
 %   \begin{minipage}[t]{\columnwidth}\centering
    \begin{tabular}{l l l}
    Parameter                                                                                     & Value                             & Units                           \\
    \hline
    $\kappa_\mathrm{0}$\footnote[1]{from Ref.~\cite{Kittel1986}}                                  & 72                                & WK$^\mathrm{-1}$m$^\mathrm{-1}$ \\
    $k_\mathrm{450nm}$\footnote[2]{from linear interpolation of data in Ref.~\cite{Johnson1974} } & 3.2540                            & -                               \\
    $C_1$\footnote[3]{from fit in Ref.~\cite{Szymanski2007}}                                      & -3.577785$\times$10$^\mathrm{-5}$ & JK$^\mathrm{-1}$m$^\mathrm{-3}$ \\
    $C_2$\footnotemark[3]                                                                         & 2.801128$\times$10$^\mathrm{6}$   & JK$^\mathrm{-1}$m$^\mathrm{-3}$ \\
    $T_0$\footnotemark[3]                                                                         & 62.98191                          & K                               \\
    $p$\footnotemark[3]                                                                           & 2.06271                           & -                               \\
    $m$\footnotemark[3]                                                                           & 278.44328                         & JK$^\mathrm{-2}$m$^\mathrm{-3}$ 
    \end{tabular}
    \end{ruledtabular}
 %   \end{minipage}
    \label{tab:TTM}
\end{table}
We have considered across all our TTM calculations that $C_\mathrm{e}$ and $G$ are explicitly dependent on the temperature of the electron bath subsystem. Unlike for $T_\mathrm{l}$, the maximum $T_\mathrm{e}$ value that can be reached during the intense but short laser pulse irradiation is well above the melting point of the metal substrate, for which it is very difficult to obtain experimentally. Therefore, it is typical of TTM studies to model any $T_\mathrm{e}$ dependent quantity by extrapolation of fitted data at relatively low temperatures to higher electronic temperatures or by using theories valid in the limit $T\rightarrow 0$ for simplicity. In order to avoid this problem, we have taken advantage of the recently calculated theoretical temperature dependence of $G$ and $C_\mathrm{e}$ in palladium by Li and Ji~\cite{Li2022} (Figs.~4 and 5 therein), which covers temperatures from 300~K to 20\,000~K. In the latter work, both $G$ and $C_\mathrm{e}$ are calculated from first principles as explicitly dependent on the electron DOS and Fermi-Dirac distribution of electrons in palladium, which are allowed to change with electronic temperature.

In Fig.~\ref{fig:CeG}, we compare several models of $C_\mathrm{e}(T_\mathrm{e})$ (left panel) and $G(T_\mathrm{e})$ (right panel) found in the literature with the values calculated by Li and Ji~\cite{Li2022} (red full lines). With a blue dotted line (left panel) we plot $\gamma T_\mathrm{e}$, a linear model of $C_\mathrm{e}$ fitted to experiments at very low temperatures ($T_\mathrm{e}<$50~K)~\cite{Miiller1971}. This is the standard theoretical behavior of $C_\mathrm{e}$ derived from Drude model and from a free electron gas~\cite{Ashcroft1988}. The green-dashed lines $C_\mathrm{e}^{\mathrm{T}\downarrow}$ (left panel) and $G^{\mathrm{T}\downarrow}$ (right panel) correspond to the $C_\mathrm{e}$ and $G$ models used in prior laser induced desorption studies on palladium~\cite{Szymanski2007,Hong2016}. The superscript $\mathrm{T}\downarrow$ denotes that these results were obtained at low temperatures and extrapolated for higher values of $T_\mathrm{e}$. $C_\mathrm{e}^{\mathrm{T}\downarrow}$ was calculated as a linear fit $\gamma_\mathrm{fit}T+b$ to experimental data measured below 1000~K in an effort to maintain the linear behavior of $C_e$ predicted by simple theoretical models while still getting numerical predictions close to experiments. $G^{\mathrm{T}\downarrow}$ was obtained from the phonon spectrum second moment measured in Ref.~\cite{Miiller1971} at 296~K and kept constant with respect to $T_\mathrm{e}$. Figure~\ref{fig:CeG} shows that the $C_\mathrm{e}$ and $G$ values calculated by Li and Ji lie below their counterparts extrapolated from low temperature measurements. In the case of the electronic heat capacities (left panel), $C_\mathrm{e}$ and $C_\mathbf{e}^{\mathrm{T}\downarrow}$ are very close for temperatures below 2000~K, point at which both quantities start to diverge up to a factor of two when $T_\mathrm{e}=$8000~K. The linear extrapolation $\gamma T_\mathrm{e}$, typical of a free electron gas, is clearly overestimating the electronic heat capacity even at relatively low temperatures, which states that this approximation is not good to model laser-palladium interaction in the context of our work. In the case of $G$ and $G^{\mathrm{T}\downarrow}$ (right panel), there is a fast decrement of $G$ with $T_\mathrm{e}$ that is completely lost by the constant behavior of $G^{\mathrm{T}\downarrow}$ making it overestimating the energy flow exchange between electrons and phonons by a factor higher than three when $T_\mathrm{e}>$2500~K.

\acknowledgments{ The authors acknowledge financial support by 
the Gobierno Vasco-UPV/EHU Project No.  IT1569-22, 
Generalitat de Catalunya grant 2021 SGR 01411,
and the Spanish Ministerio de Ciencia e Innovación [Grants No. PID2019-107396GB-I00 and No. PID2020-113565GB-C21 MCIN/AEI/10.13039/501100011033]. 
D.N. additionally acknowledges financial support from the Croatian Science Foundation (Grant no. UIP-2019-04-6869).
This research was conducted in the scope of the Transnational Common Laboratory (LTC) “QuantumChemPhys – Theoretical Chemistry and Physics at the Quantum Scale”. Computational resources were provided by the DIPC computing center. R.B. acknowledges European  Union-NextGenerationEU, Ministry of Universities and Recovery, Transformation and Resilience Plan, through a call from Polytechnic University of Catalonia.
}
\bibliography{refs}

\end{document}